\documentclass[12pt,a4paper]{ouparticle} 
\newcommand\rcsinfo{\$Revision: 1.55.1.3 $ $ -- \$Date: 2016/04/29 20:36:32 $ $ }

\usepackage{multirow}

\usepackage{mathtools}
\usepackage{graphicx}

\usepackage{amsmath} 
\usepackage{natbib}
\bibpunct{(}{)}{;}{a}{}{,}
\usepackage[usenames,dvipsnames]{color}
\usepackage{url}
\usepackage{bibunits}
\defaultbibliography{zbb-abbrev}
\defaultbibliographystyle{elsart-harv}

\usepackage[table,rgb]{xcolor}
\usepackage{colortbl}
\usepackage{tikz}

\usepackage{version}
\excludeversion{draft}
\includeversion{submit}

\usepackage{float}
\floatstyle{ruled}
\newfloat{floatbox1}{t}{lob}
\floatname{floatbox1}{Box}

\usepackage{algorithm2e}

\usepackage[running]{lineno}
\usepackage{patchcmd}  
\patchcommand\align{\hrule height 0pt depth 0pt width 0pt}{}
\patchcommand\multline{\hrule height 0pt depth 0pt width 0pt}{}

\usepackage{multirow}

\newcommand{\PTCx}{Pressure (`$F$') Target Control}
\newcommand{\STCx}{State (`$B$') Target Control}
\newcommand{\PTC}{Pressure Target Control}
\newcommand{\STC}{State Target Control}
\newcommand{\SOC}{Self-Optimising Control}
\newcommand{\CFP}{Single Species Control} 
\newcommand{\NP}{Nash Pressure}
\newcommand{\NS}{Nash State} \newcommand{\TY}{Total Yield}
\newcommand{\OM}{SM}
\newcommand{\tr}{^{\scriptscriptstyle\mathrm T}}
\newcommand{\diag}{\mathrm{diag}}

% ...

\newlength{\figurewidth}
\begin{submit}
  \makeatletter
  \@input@{supp-mat.aux}
  \makeatother
\end{submit}

\begin{document}

\begin{draft}
  \title{\bf Maximum sustainable yield from interacting fish stocks in an
  uncertain world: two policy choices and underlying
  trade-offs\newline\footnotesize\rcsinfo}
\end{draft}
\begin{submit}
  \title{\bf Maximum sustainable yield from interacting fish stocks in an
  uncertain world: two policy choices and underlying
  trade-offs}
\end{submit}

\author{%
\name{Adrian Farcas}
\address{Centre for Environment, Fisheries \& Aquaculture Science\\
  Pakefield Road, Lowestoft NR33 0HT,
  United Kingdom}
\email{adrian.farcas@cefas.co.uk}
\and
\name{Axel G. Rossberg\thanks{corresponding author}}
\address{Queen Mary University of London, School of Biological and
  Chemical Sciences, 327 Mile End Rd, London E1, United Kingdom
  \newline and \newline
  Centre for Environment, Fisheries \& Aquaculture Science\\
  Pakefield Road, Lowestoft NR33 0HT,
  United Kingdom}
\email{a.rossberg@qmul.ac.uk}
}

\begin{draft}
  \date{\small Draft, \today\\\tiny\rcsinfo}
\end{draft}
\begin{submit}
  \date{\vspace*{-0.1cm}26 May 2016\newline\small \copyright Crown copyright}
\end{submit}
\abstract{%
  The case of fisheries management illustrates how the inherent
  structural instability of ecosystems can have deep-running policy
  implications.  We contrast ten types of management plans to achieve
  maximum sustainable yields (MSY) from multiple stocks and compare
  their effectiveness based on a management strategy evaluation (MSE)
  that uses complex food webs in its operating model.  Plans that
  target specific stock sizes ($B_{\text{MSY}}$) consistently led to
  higher yields than plans targeting specific fishing pressures
  ($F_{\text{MSY}}$).  A new self-optimising control rule, introduced
  here for its robustness to structural instability, led to
  intermediate yields.  Most plans outperformed single-species
  management plans with pressure targets set without considering
  multispecies interactions.  However, more refined plans to
  ``maximise the yield from each stock separately'', in the sense of a
  Nash equilibrium, produced total yields comparable to plans aiming
  to maximise total harvested biomass, and were more robust to
  structural instability.  Our analyses highlight trade-offs between
  yields, amenability to negotiations, pressures on biodiversity, and
  continuity with current approaches in the European context.  Based
  on these results, we recommend directions for developments of EU
  fisheries policy.}


\maketitle

\begin{bibunit}

\section{Introduction}
\label{sec:introduction}

\subsection{Four questions about multispecies MSY}
\label{4questions}

Fisheries management aiming to attain maximum sustainable yield (MSY)
from multiple interacting stocks is considerably more
complicated than single-stock management
\citep{pope76:_mixed_fish,bulgakova86:_analy_cape_cape_icseaf_div,
  collie03:_using_amoeb, Walters05:_NashPressure,
  Matsuda06:_MaximalYields, Gecek12:_ImpactMaximumMSY,
  Houle13:_SizeSpectrumMSY,voss14:_region_mmsy,
  Thorpe16:_AssessingFishery}.  \textit{A
  priori} it is not even clear what the best translation of the MSY
objective for a single, isolated stock is to cases with multispecies
interactions, i.e.\ with feeding and competitive interactions among
species 
(for simplicity, we use ``stock'' interchangeable with ``fish species'' in this paper). 
Even when a
management objective 
considering
multispecies interactions is
defined, attaining it can be difficult because these interactions are
generally not well known. It is therefore not surprising if
legislation aiming at MSY acknowledges the role of multispecies
interactions in principle, but tends to play it down.  Examples are
\S\S301.a.3, 303.b.12 of the Magnuson-Stevens Act
\citep{commerce07:_magnus_steven_act} or 
Article 9.3.b of the Common Fisheries Policy \citep{eu13:_cfp}, but
see, \S13(2) of New Zealand's Fisheries Act
\citep{office14:_fisher_act2}.  In the current practice of fisheries
management, multispecies interactions among managed stocks play only a
minor role.  Noteworthy exceptions are cases where stock assessments
take account of the dependence of predation mortality on the
abundances of other species
\citep{Gjosaeter15:_RetrospectiveEvaluation, ices14:_wgsam}.

To contribute to the development of management practices mindful of
multispecies interactions we ask here:
\begin{enumerate}
\item[Q1:] How can the singe-species MSY objective be translated into
  the multispecies case?\label{item:objectives}
\item[Q2:] Which strategies are suited to achieve such
  objectives?\label{item:rules}
\item[Q3:] How do these options compare with regards to the yields
  they achieve, the degree of collaboration among players required to
  reach the objectives, pressures on biodiversity, and their political
  acceptability?
  \label{item:outcomes}
\item[Q4:] How much can be gained from multispecies management in
  comparison with management disregarding ecological interactions?\label{item:gain}
\end{enumerate}

Three conceivable answers to each Q1 and Q2 are given in the following
two sections.  Hence there are two high-level policy choices to be
made.  The first choice is of the management \emph{objective}, the
second of the type of \emph{strategy} employed to achieve it.  Each of
the resulting nine variants we call a \emph{management plan}.

To answer Q3, we performed formal management strategy evaluations
(MSE, \citealt{walters76:_adapt,hilborn92:_quant}) of these plans.
That is, the outcomes of the plans were evaluated by applying them to
fisheries in hypothetical ecosystems, described by an \emph{operating
  model}, that are not fully known to the manager.  To answer Q4,
evaluation scores were compared with those for management that aims at
MSY for the same set of stocks while disregarding multispecies
interactions, modelled after current EU practice.

\subsection{First policy choice: the management objective}
\label{sec:choice-objectives}

There is no unique way of translating the single-species MSY objective
to the multispecies case.  Maximisation of yield from one stock will
generally require different strategies than maximisation of yield from
another.  In a simple predator-prey system, for example, the
maximisation of yield from the prey requires culling the predator,
while by not exploiting the prey yield from the predator can be
maximised \citep{may79:_msy, clark90:_bioec}.

The three types of high-level objective we consider here are:
\begin{itemize}
\item[-] \emph{Nash Pressure}: To fish all exploited stocks at such rates
  that changes in the exploitation rate of any single stock cannot
  increase the long-term yield from that stock.
\item[-] \emph{Nash State}: To keep all exploited stocks at such sizes
  (e.g.\ in terms of spawning stock biomass) that changes in the size
  of any single stock cannot increase the long-term yield from that
  stock.
\item[-] \emph{Total Yield}: Maximisation of the summed long-term yield
  from all exploited stocks.
\end{itemize}
These objectives are generally equivalent only in absence of
multispecies interactions.

\begin{figure}
  \begin{center}
    \includegraphics[width=0.65\textwidth]{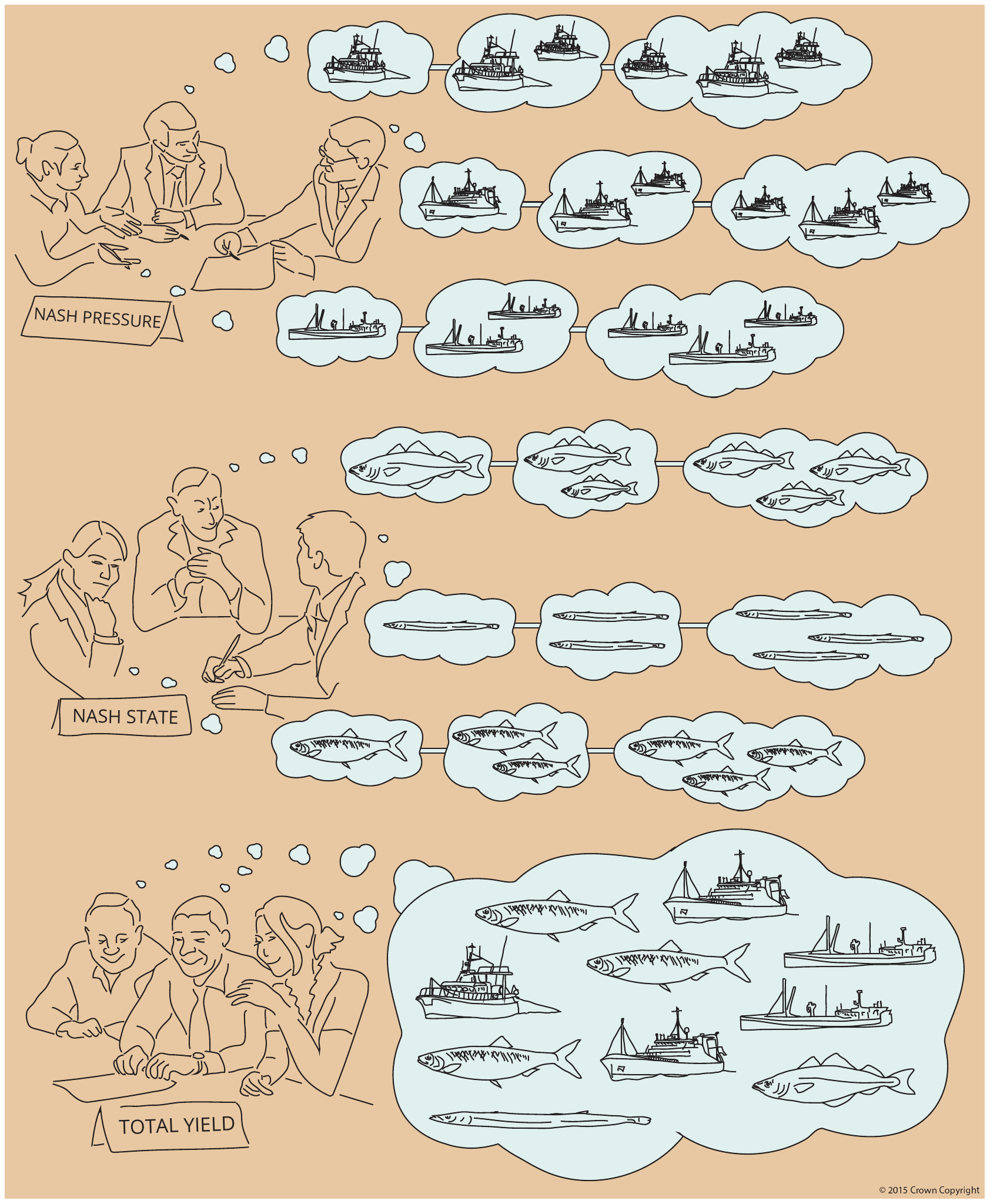}
  \end{center}
  \caption{Illustration of the first high-level policy choice as an
    option between three hypothetical games. The games correspond to
    the three high-level management objectives we consider: Nash
    Pressure, Nash State and Total Yield. The Nash Pressure players
    are managers of hypothetical fleets targeting a single stock, each
    aiming to maximise yield from that stock by adjusting its
    exploitation rate. The Nash State players are hypothetical
    managers of individual stocks, aiming to maximise yield from that
    stock by adjusting its equilibrium abundance. The last game,
    maximisation of total yield, requires collaboration between all
    players.  Infographic design by Georgia Bayliss-Brown.}
  \label{infographic}
\end{figure}

The objectives Nash Pressure and Nash State are two ways of
implementing the idea of ``maximisation of yield from each stock
separately''.  They correspond to the \emph{Nash equilibrium} outcome
in game theory \citep{osborne94:_game_theory}, where no player of a
game could improve its gains by changing its moves.  Nash equilibria
are traditionally understood as arising naturally when players are not
collaborating.  For the Nash Pressure objective the players are
hypothetical fleets, each targeting one specific stock, and the
permitted moves changes in their exploitation rates.  For the Nash
State objective the players are hypothetical managers of individual
stocks and their moves are the stock sizes they aim at.  The Total
Yield objective corresponds a situation where players chose their
moves such that the total gain by all players is maximised.  Attaining
this objective generally requires collaboration or enforcement through
governing institutions.  Figure~\ref{infographic} depicts these three
hypothetical games and objectives.

Variants of \TY{}
\citep{pope76:_mixed_fish,bulgakova86:_analy_cape_cape_icseaf_div,
  Matsuda06:_MaximalYields, Gecek12:_ImpactMaximumMSY,
  Houle13:_SizeSpectrumMSY,voss14:_region_mmsy} and \NP{} objective
(\citealt{Walters05:_NashPressure, collie03:_using_amoeb}; Moffit et
al., 10.1016/j.dsr2.2015.08.002, in press) have been considered in the
literature.  We are unaware of studies addressing the \NS{} objective.

\subsection{Second policy choice: the strategy}
\label{sec:choice-hcr}

The management strategies that we consider differ by the structure of
the corresponding harvest control rules \citep{Deroba08:_ReviewHCR}.
All rules ultimately prescribe the total catch from each stock for a
given time period, e.g.\ a year.  This corresponds to total allowable
catches (TAC) if one assumes that allowances are fully used.  Combined
with estimates of stock sizes, rules for catches can be formulated in
terms of rules for fishing mortality rates.  Depending on how these
rates are determined from managers' knowledge of stock sizes and
interactions, we consider:
\begin{itemize}
\item[-] \emph{\PTC}, where fishing mortality rates are kept fixed
  at target values deemed to be consistent with the
  selected objective.
\item[-] \emph{\STC}, where fishing mortalities are continuously
  modified such as to adjust stock sizes to target values deemed to be
  consistent with the selected objective.
\item[-] \emph{\SOC}, where neither targets for fishing mortalities
  nor stock sizes are fixed, and instead target fishing mortalities
  are varied according to homogeneous linear functions of stock sizes.
  These functions are chosen such that the expected resulting
  equilibrium state is consistent with the selected objective.
\end{itemize}
Formal mathematical definitions of these objectives are given in
Supplementary Material, \OM{}~\ref{sec:msy-pdmm}.


\subsection{Simplifications, complications, negotiations}
\label{sec:complications}

To simplify analysis and discussion, and to isolate the particular
implications of multispecies interactions, a number of complications
relevant in the practice of management are not covered in our
analysis.  First, plans do not generally include constraints resulting
from conservation objectives.  There are several conceivable
implementations of such constraints, among others those associated
with $F_{\text{PA}}$, $B_{\text{PA}}$, or $B_{\text{trigger}}$
reference points \citep{ices14:_gener_ices, multi_plans14:_final}, and
we would not want to confound the side effects of specific
implementations with the implications of the policy choices studied
here.  Instead, we determine biodiversity impacts of the unconstrained
management plants.  Second, neither technical interactions among
fleets, that is, complications due to limitations to stock-selectivity
of fishing, nor the resulting issues surrounding by-catch and discards
are considered.  Third, we disregard differences in market value of
species and other factors differentiating between biomass and economic
yields.  Forth, environmental stochasticity, including recruitment
variability, is not considered, despite its importance for stock
assessments and decision making.  Real-world management must take all
these and many other complications into account and adapt objectives
and strategies accordingly.  The plans we consider could be modified
to this end, at moderate cost in complexity and presumably with little
effect on the trends we find.

However, managing each of these complications involves value
judgements, and presumably requires negotiations among stakeholders.
Rather than prejudicating such negotiations, one can ask how easy the
negotiations would be when based on the different objectives we
investigate.  By definition, long-term yield is maximal for each
player under the Nash \NP{} objective, and so will not decline much
when the player fishes at slightly different rates---and similarly for
\NS{} (impacts on yields from other stocks, however, can be larger!).
Stakeholders will therefore be open to negotiations of their targets.
This intuition is captured by proposals to define MSY as a ``range''
of targets rather than a particular ``point'' (Task Force on
Multiannual Plans \citeyear{multi_plans14:_final}), and to use this
flexibility to take technical constraints into account.  Achievement
of the \TY{} objective generally involves trading the yield from one
stock off against that from another, generating winners and losers
among fleets \citep{voss14:_region_mmsy}.  Negotiations aiming for
\TY{} can therefore be more difficult.  The question arises what the
loss in yields would be when opting for a Nash equilibrium to avoid
these difficulties. This is addressed below.

As a technical simplification, required for computational reasons, our
models do not structure populations by size or age.  For food-web
interactions, corresponding population dynamics can be derived from
size-structured models using projection methods
\citep{rossberg11:_qna, rossberg12:_analytic_size_spectrum}.  However,
this means we disregard size-selectivity effects.  We consider this a
legitimate simplification, because, in a first approximation,
optimisation of size selectivity and of fishing mortality for given
size selectivity are separate problems \citep{law88:_msy_and_repval,
  getz12:_harves_theor, scott11:_long_term_yield}.

\subsection{Structural instability of natural and modelled ecological
  communities}

A particular concern of this study are the management
implications of structural instability of ecological communities
\citep[Chapter~18]{rossberg13:_food_webs_biodiv}.  By \emph{structural
  instability} we mean that, in a community of interacting
populations, 
small changes in environmental conditions or external pressures can
lead to large changes in population sizes
\citep{yodzis88:_structural_instability}.  Structural instability
manifests itself, among others, through the difficulties fisheries
modellers experience in parameterizing models of interacting species
such that these reproduce observed community structure and
dynamics. Small deviations of model structure or parameters from
reality can lead to very different model states.  While the phenomenon
as such is well known
\citep{andersen77:_community_model, flora11:_struc_sens,
  hill07:_model, novak11:_predic_responses,
  yodzis88:_structural_instability}, the likely underlying ecological
mechanism has only recently been revealed
\citep[Chapters~14-18]{rossberg13:_food_webs_biodiv}. Because
structural instability increases with species richness
\citep{novak11:_predic_responses}, naturally assembled communities
eventually saturate at richness levels where structural instability is
so large that the perturbation from one species' invasion leads, on
average, to extirpation of one other species.

Structural instability poses a dilemma to every fisheries modeller
interested in the implications of multispecies interactions.  If
models have less structural instability than in nature, pressure-state
relations might not be correctly represented, leading to incorrect
projections for MSY.  If, on the other hand, the model's structural
instability is as large as in nature, the model will be difficult to
parameterize.  Besides, comparing a model's structural instability to
reality is difficult.  In our MSE, these problems are avoided by
abandoning the idea of using an operating model representative of a
specific natural community---for our general analysis, this might not
even be desirable.  Instead, model communities are constructed through
a random process that mimics natural assembly and turnover of aquatic
food webs.  This leads to the kind of community saturation thought to
be responsible for structural instability in nature
\citep{Borrelli15:_SelectionStability}.  Since, in addition, the
resulting communities have macroecological properties similar to
marine communities \citep{fung13:_why_slow_recover}, we expect them to
exhibit structural instability of a degree similar to that of marine
systems.

We introduce below a method (``conservatism'') capable of mitigating
the adverse effects of structural instability on management outcomes.
In essence, it consists in avoiding management targets that are too
different from the current community state to be reliably described by
inherently inaccurate management models.  Instead, management
gradually adapts targets to meet the objective.

It turns out that the answers to above questions Q3 and Q4 are
essentially determined by structural instability.  Key results of our
MSE can be anticipated mathematically from structural instability and
general principles of community dynamics alone, as we highlight below
by referring to corresponding mathematical considerations developed in
Supplementary Material.  The good agreement between general
theoretical expectations and simulations using a complex model suggests
that these expectations are sufficiently robust to be fulfilled in the
real world as well. This justifies us in deriving recommendations for
practical fisheries management from this study.

\bigskip



\section{Methods}
\label{sec:methods-summary}

\subsection{Operating model}
\label{sec:operating-model}

The operating model representing ``reality'' in our MSE is the
Population Dynamical Matching Model \citep[PDMM;][]{rossberg07:_flat,
  fung13:_why_slow_recover,rossberg13:_food_webs_biodiv}.  It
describes structure and dynamics of aquatic food webs with species
resolution.  Species in the PDMM have different maturation body
masses, which determine, through allometric scaling laws
\citep{yodzis92:_body_size_dyna}, maximum growth rates and
consumption-independent loss rates (metabolic losses and
non-predation mortality). The consumption of a resource species by a
consumer species is modelled through a Type II functional response,
modified to model prey-switching following
\cite{leeuwen13:_switching}.  For any two species in the PDMM, their
trophic interaction strength depends on their relative body masses and
the match between two kinds of abstract traits that characterise them
as consumers and resources. Size preference is parameterized through a
population-level predator-prey size ratio window
\citep{rossberg12:_analytic_size_spectrum}.  For a full description of
the model, see \citeauthor{rossberg13:_food_webs_biodiv}
(\citeyear{rossberg13:_food_webs_biodiv}, Chapter~22) and
\OM~\ref{sec:popul-dynam-match}.

In the parameterization used here, the model spans approximately five
trophic levels and resolves species and their interactions over $17$
orders of magnitude in maturation body size (median range $10^{-15}$
to $10^{1.8}\,\mathrm{kg}$). Following
\cite{fung13:_why_slow_recover}, species with maturation body sizes
larger than $1\,\mathrm{g}$ are interpreted as fish.  When numbering
the $S$ species in the model from the largest to the smallest, so the
first $S_F$ species are the fish, this leads to population dynamics of
the general form
\begin{subequations}
  \label{eq:pdmm-general}
  \begin{align}
    \label{eq:pdmm-general-fish}
    \frac{dB_i}{dt} &= g_i(B_1,\dots,B_S)B_i-F_iB_i&&(1\le i\le
    S_F),\\
    \label{eq:pdmm-general-other}
    \frac{dB_i}{dt} &= g_i(B_1,\dots,B_S)B_i&&(S_F< i\le S),
  \end{align}
\end{subequations}
where $t$ is time.  The momentary growth rate $g_i(\dots)$ of the
population biomass $B_i$ of each species $i$ depends on its direct
biological interactions with other species.  The parameters $F_i$
represent exploitation rates, and for each fish species $i$ the
product $Y_i=F_iB_i$ is the yield from that species per unit time.
Exploitation rates are proportional to adult fishing mortalities and
typically attain numerically similar values
\citep{shephard12:_size_celtic_sea}, hence our use of the symbol $F$.

PDMM model communities are generated by simulating the natural
processes of community assembly and turnover by iteratively adding
species to the model and removing those that go extinct, until $S$
fluctuates around some equilibrium.  This is done with all
exploitation rates ($F_i$) set to zero.  Each MSE was performed using
$37$ different samples from this community, taken after every $10.000$
species additions and therefore composed of largely independent sets
of species (of $40$ communities so sampled, the first $3$ were
discarded as burn-in). In our parameterization
(\OM~\ref{sec:popul-dynam-match}), each sample contained around $2000$
species, of which between $9$ and $38$ were fish (mean $22.4$).

\subsection{Management model}
\label{sec:management-model}

As the management model, we used the multispecies extension of
Schaefer's (\citeyear{schaefer54:_some}) surplus production (or
multispecies Lotka-Volterra) model \citep{pope76:_mixed_fish,
  pope79:_stock_gulf_thail}
\begin{equation}
  \label{eq:surplus-production}
  \frac{dB_i}{dt}=\left[r_i-\sum_{j=1}^{S_F}G_{ij}B_j\right]B_i-F_iB_i,
\end{equation}
where the surplus productivities $r_i$ and interaction coefficients
$G_{ij}$ ($1\le i,j\le S_F$) are constants.  This model has
frequently been studied for management applications
\citep{bulgakova86:_analy_cape_cape_icseaf_div, collie03:_using_amoeb,
  gaichas12:_assem, ices89:_repor_multis_asses_workin_group,
  pope79:_stock_gulf_thail}.  It has the advantages of compatibility
with the abstractions employed in the operating model,
Eq.~\eqref{eq:pdmm-general}, formal simplicity, and a low number of
fitting parameters compared to processed-based models.  Management
models used in practice are often more complex.

\OM~\ref{sec:calibr-mang-model} describes a method to
calibrate the parameters of the management model ($r_i$, $G_{ij}$,
$1\le i,j\le S_F$) such that it approximates the dynamics of fish
populations in the operating model, Eq.~\eqref{eq:pdmm-general}, for
states similar to the current.  The method is based on the assumption
that the responses of non-fish species to changes in fish populations
are fast compared to these changes.  It should therefore work best for
fish populations close to equilibrium, where they change slowly.

A mathematical analysis of the calibration method reveals that
structural instability of the operating model can lead to inaccurate
approximations of the operating model by the management model and to
structural instability of the management model itself (\OM~\ref{sec:effects-struct-inst-mod}).  Our MSE is designed to
capture the implications for management outcomes of these effects and
of incomplete representation of reality by the management model,
because similar issues can arise for management models used in
practice.  For simplicity, the MSE does not consider measurement
errors of population-dynamical parameters or implementation errors,
and it assumes perfect knowledge of all $B_i$ and and $F_i$.

Management plans are applied in cycles simulating adaptive management.
At the beginning of each cycle, the management
model 
is calibrated to approximate the dynamics of the operating
model 
for states similar to the current state.  The multispecies harvest
control rule (mHCR, see below) corresponding to the plan is then
(re)parameterized using the management model and used throughout the
management cycle.  A rather long period ($50$ years) is chosen for
these cycles to allow the operating model to reach an equilibrium,
because, with our simple calibration algorithm, this results in better
fit of the management model to the operating model in the next cycle.
Conceivable effects of shorter management periods were not
investigated.

\subsection{Multispecies harvest control rules}
\label{sec:mult-harv-contr}

The three management strategies we consider differ by the formulae
used to determine exploitation rates (mHCR). The free parameters in
these formulae ($F_{\text{MSY},i}$, $B_{\text{MSY},i}$, $\hat{G}_i$)
are chosen such that, if the management model was the correct model of
reality, the corresponding equilibrium state of the system would meet
the objective of the plan.  Corresponding analytic expressions are
derived in \OM~\ref{sec:solv-oper-model} following
\cite{pope79:_stock_gulf_thail}.

The simplest case is \PTC{}, where
exploitation rates are kept constant at
\begin{align}
  \label{eq:PTC}
  F_i&=F_{\text{MSY},i}&&\text{(\PTC)}
\end{align}
for each fish species $i$.

\STC{} is implemented by a rule
\begin{align}
  \label{eq:STC}
  F_i&=\max\left[0,F_{0,i}+\frac{1}{T}\left(1-\frac{B_{\text{MSY},i}}{B_i}\right)\right]&&\text{(\STC)},
\end{align}
where the $F_{0,i}$ are defined in the following paragraph,
$B_{\text{MSY},i}$ are the targeted stock sizes, and the parameter $T$
(dimension Time) depends on how fast management attempts to reach the
target.  Here we chose $T=1\,\text{yr}$ rather small, to model ``fixed
escapement'' management as recommend by bio-economic analyses with
low discount rate.  The operation $\max\left[0,\ldots\right]$ replaces
any negative exploitation rate by zero.

As the (time dependent) \emph{neutralising exploitation rate}
$F_{0,i}$ we define the exploitation rate that would keep stock $i$ at
its current size, provided all other stocks and long-term
environmental conditions remain in their current state as well.  The
values $F_{0,i}$ can be computed independently of the management model
Eq.~\eqref{eq:surplus-production}, in practice during yearly
stock-assessments using standard methods, or their multispecies
extensions.  In our MSE we set $F_{0,i}=g_i(B_1,\ldots,B_S)$, with
$g_i(B_1,\ldots,B_S)$ as in Eq.~\eqref{eq:pdmm-general-fish}.  As a
result, Eq.~\eqref{eq:pdmm-general-fish} become effectively
\begin{align}
  \label{eq:effective-dynamics}
  \frac{dB_i}{dt}=\frac{1}{T}\left(B_{\text{MSY},i}-B_i\right) &&&
  \text{(effective dynamics under \STC{})}
\end{align}
whenever $F_i$, as given by Eq.~\eqref{eq:STC}, is positive.  That is,
the \STC{} rule is designed to achieve exponential approach of all
stocks to their target sizes $B_{\text{MSY},i}$ at a rate $1/T$.  By
modifying Eq.~\eqref{eq:STC} and/or the parameter $T$, other dynamics
for approaching $B_{\text{MSY},i}$ can be obtained.  For the type of
plan we recommend in Conclusions below, one such variant will briefly
be discussed.

For the \NP{} and \NS{} objective, the \SOC{} rule 
has the form
\begin{align}
  \label{eq:SOC}
  F_i&=\max\left[0,\hat{G}_i B_i\right] &&&\text{(\SOC{} I)},
\end{align}
where the constants $\hat{G}_i$ depend on the interaction coefficients
$G_{ij}$ of the fitted management model,
Eq.~\eqref{eq:surplus-production}.  On average, approximately $4\%$ of
the calculated $\hat{G}_i$ were negative for both \NP{} and \NS{} and
therefore $F_i$ set to zero.

\SOC{} for the \TY{} objective requires 
\begin{align}
  \label{eq:SOC-TY}
  F_i=\max\left[0,\sum_{j=1}^{S_F} G_{ji} B_j\right]&&&\text{(\SOC{} II)}.
\end{align}
The coefficients $G_{ji}$ entering Eq.~\eqref{eq:SOC-TY} are directly
given by interaction strengths from the fitted surplus production
model, Eq.~\eqref{eq:surplus-production}.
Because the matrix $(G_{ji})$ entering Eq.~\eqref{eq:SOC-TY} is the
transpose of the interaction matrix $\mathbf{G}=(G_{ij})$ from
Eq.~\eqref{eq:surplus-production}, we call the scheme in
Eq.~\eqref{eq:SOC-TY} \emph{Transposed Interaction Control}, alluding
to the ``Transposed Jacobian'' scheme of control theory
\citep{craig89:_introd}.

%

\subsection{Single species harvest control rule}
\label{sec:single-sp-hcr}

To model \CFP{} as currently being implemented in the EU
\citep{multi_plans14:_final}, we fixed target exploitation rates as in
Eq.~\eqref{eq:PTC}, but now computed the targets
$F_{\text{MSY},i}=F_{\text{SSC},i}$ using single-species surplus
production models, fitted to the operating model's apparent dynamics
for that species when disregarding all others.  See
\OM~\ref{sec:cfp-parameters}.

\subsection{Addressing structural instability through conservative management plans}
\label{sec:struct-conserv}

Mathematical analyses suggest that, due to structural instability, the
parameters of mHCRs can sensitively respond to small errors in the
population-dynamical parameters of management models (\OM~\ref{sec:effects-struct-inst-par}) and that inaccuracies in
mHCR parameters can lead to large reductions in yields
(\OM~\ref{sec:effect-struct-refpoints}).

It might therefore be advantageous to pursue management strategies
that are conservative in the sense of preferring targets that differ
less from current states than other, non-conservative strategies, even
if the latter are predicted to be optimal by a naive evaluation of the
management model.  We achieve such conservatism through modifications
of the formulae for mHCR parameters. Technically, we implement a
variant of Lavrentiev regularisation of matrix inversion
\citep{engl2000:_regularization}, which improves the numerical
robustness of the calculation, see \OM~\ref{sec:impl-cons-reg}. A
regularisation parameter controls the degree of conservatism.  Where
applicable, we performed the MSE both with that parameter set to zero
(\emph{no conservatism}) and to a fixed reasonable non-zero value,
given in \OM~\ref{ssec:reg} (\emph{standard conservatism}).  To obtain
upper bounds on conceivable further improvements of outcomes through
conservatism, we evaluated in addition plans for the case where the
parameter was chosen for each plan and sample community such that the
actual total long-term yield attained was maximised (\emph{optimal
  conservatism}).

\subsection{Theoretical maximum sustainable total yield}
\label{sec:theor-maxi-total}

As a yardstick to compare the outcomes of management plans against, we
computed the theoretical maximum long-term total yield
($\text{MSY}_{\text{theo}}$) achievable from each sampled model
community.  This was determined using an evolutionary optimisation
algorithm \citep{Stafford08:_ComputationalApproachMSY}.  Specifically,
we applied the method of \cite{runarsson05:_searc} as implemented in
\texttt{NLopt} (\url{http://ab-initio.mit.edu/wiki/index.php/NLopt}),
modified to improve convergence (\OM~\ref{sec:modif-evol-search}).

\subsection{Comparison of management plans}
\label{sec:comp-manag-strat}

We compared all possible management plans, i.e., combinations of
objectives and strategies, based on the total yield they achieved and
the resulting impacts on biodiversity and community size structure.
The long-term yield resulting from applying a plan to a sample
community was computed as the time-averaged yield over the duration of
$8$ adaptive management cycles, after a $24$ cycle transition phase.
Because $\text{MSY}_{\text{theo}}$ varied substantially among sample
communities ($\text{CV}=0.40$), we quantified the yield from each plan
as the average over all sample communities of the percentage of
$\text{MSY}_{\text{theo}}$ attained.

As secondary criteria for comparison, we computed the proportion of
fish species conserved ($100\%-\text{proportion extirpated}$) and the
mean logarithmic maturation body mass of species at the end of each
MSE run.  The latter addresses the specific concern that maximisation
of yields might requires culling large piscivorous species to release
pressure on smaller and more productive planktivores.  As for yields,
both indices were averaged over all sample communities.  For
comparison, we calculated these indices also for unfished communities
after simulating them over a period corresponding to $32$ management
cycles, because slow residual dynamics of stocks
\citep{fung13:_why_slow_recover} can result in natural extirpations.

In the policy context, other criteria will also play a role, for
example the similarity of plans to those currently in place or, based
on game theoretical considerations, the likelihood with which
stakeholders will agree to plans or small modifications of them.
Rather than attempting to score plans based on such criteria, key
considerations are suggested in Discussion.

\section{Results}
\label{sec:results}

\begin{table}
  \caption{Illustration of trade-offs among multispecies management plans to
    achieve MSY, as derived from our MSE.  Intensity of colouring increases with
    score attained.\newline}
  \label{tab:results}

  \small
  \centering
\newcommand{\boxcolor}[1]{\pgfmathparse{(1-(#1-0.3))}\xglobal\definecolor{mC}{rgb}{1,\pgfmathresult,\pgfmathresult}\cellcolor{mC}\color{black}\pgfmathparse{100*#1}\pgfmathprintnumber[precision=1,fixed zerofill=true,assume math mode=true]{\pgfmathresult}}

\newcommand{\boxcolorg}[1]{\pgfmathparse{(1-1.4*(#1-0.4))}\xglobal\definecolor{mC}{rgb}{\pgfmathresult,1,\pgfmathresult}\cellcolor{mC}\color{black}\pgfmathparse{100*#1}\pgfmathprintnumber[precision=1,fixed zerofill=true,assume math mode=true]{\pgfmathresult}}

\newcommand{\boxcolorb}[1]{\pgfmathparse{(1-1.4*(#1-0.5))}\xglobal\definecolor{mC}{rgb}{\pgfmathresult,\pgfmathresult,1}\cellcolor{mC}\color{yellow}\pgfmathparse{2.0102*#1}\pgfmathprintnumber[precision=2,fixed zerofill=true,assume math mode=true]{\pgfmathresult}}

\sf
\renewcommand{\arraystretch}{1.2}
\begin{tabular}{|l|l|c|c|c|c|}
  \hline
  \multicolumn{6}{|c|}{\textbf{Yield}}\tabularnewline
  \multicolumn{6}{|c|}{\textbf{(\% of theoretical maximum sustainable total yield $\text{MSY}_{\text{theo}}$)}}\tabularnewline
  \hline  
  Strategy &Conser-  & \multicolumn{4}{c|}{Objective}\tabularnewline
  \cline{3-6}
  &vatism& \multicolumn{1}{>{\centering}p{3.7em}|}{other} & \multicolumn{1}{>{\centering}p{3.7em}|}{\NP} & \multicolumn{1}{>{\centering}p{3.7em}|}{\NS} & \multicolumn{1}{>{\centering}p{3.7em}|}{\TY} 
  \tabularnewline
  \hline
  &none&&\boxcolor{0.506}&\boxcolor{0.474}&\boxcolor{0.339}\tabularnewline
  \PTCx&standard&&\boxcolor{0.567}&\boxcolor{0.382}&\boxcolor{0.561}\tabularnewline
  &optimal&&\boxcolor{0.729}&\boxcolor{0.465}&\boxcolor{0.728}\tabularnewline 
  \hline      
  &none&&\boxcolor{0.706}&\boxcolor{0.622}&\boxcolor{0.570}\tabularnewline
  \STCx&standard&&\boxcolor{0.733}&\boxcolor{0.659}&\boxcolor{0.850}\tabularnewline
  &optimal&&\boxcolor{0.834}&\boxcolor{0.732}&\boxcolor{0.930}\tabularnewline
  \hline
  \multicolumn{2}{|l|}{\SOC}&&\boxcolor{0.717}&\boxcolor{0.626}&\boxcolor{0.764}\tabularnewline
  \hline
  \multicolumn{2}{|l|}{\CFP}&\boxcolor{0.517}& & & \tabularnewline
  \hline
    \hline
    \multicolumn{6}{|c|}{\textbf{Species survival (\% initial number of fish species)}}\tabularnewline
    \hline  
    &none&&\boxcolorg{0.62}&\boxcolorg{0.733}&\boxcolorg{0.466}\tabularnewline
      \PTCx&standard&&\boxcolorg{0.689}&\boxcolorg{0.752}&\boxcolorg{0.624}\tabularnewline
      &optimal&&\boxcolorg{0.77}&\boxcolorg{0.794}&\boxcolorg{0.693}\tabularnewline
      \hline
      &none&&\boxcolorg{0.704}&\boxcolorg{0.837}&\boxcolorg{0.463}\tabularnewline
      \STCx&standard&&\boxcolorg{0.776}&\boxcolorg{0.899}&\boxcolorg{0.704}\tabularnewline
      &optimal&&\boxcolorg{0.807}&\boxcolorg{0.91}&\boxcolorg{0.727}\tabularnewline
      \hline
      \multicolumn{2}{|l|}{\SOC}&&\boxcolorg{0.889}&\boxcolorg{0.964}&\boxcolorg{0.653}\tabularnewline
      \hline
      \multicolumn{2}{|l|}{\CFP}&\boxcolorg{0.627}& & & \tabularnewline
       \hline
      \multicolumn{2}{|l|}{$\text{MSY}_{\text{theo}}$}&\boxcolorg{0.720}& & & \tabularnewline    
        \hline
       
       \multicolumn{2}{|l|}{Unfished}&\boxcolorg{0.971}& & & \tabularnewline    
    \hline    
    \hline
    \multicolumn{6}{|c|}{\textbf{Fish community mean $\log_{10}$ maturation body mass
        (grams) at the end of MSE}}\tabularnewline
    \hline  
    &none&&\boxcolorb{0.845}&\boxcolorb{0.972}&\boxcolorb{0.792}\tabularnewline
    \PTCx&standard&&\boxcolorb{0.914}&\boxcolorb{1.006}&\boxcolorb{0.886}\tabularnewline
    &optimal&&\boxcolorb{0.947}&\boxcolorb{0.987}&\boxcolorb{0.896}\tabularnewline
    \hline
    &none&&\boxcolorb{0.835}&\boxcolorb{0.926}&\boxcolorb{0.758}\tabularnewline
    \STCx&standard&&\boxcolorb{0.879}&\boxcolorb{0.976}&\boxcolorb{0.850}\tabularnewline
    &optimal&&\boxcolorb{0.897}&\boxcolorb{0.974}&\boxcolorb{0.829}\tabularnewline
    \hline
    \multicolumn{2}{|l|}{\SOC}&&\boxcolorb{0.916}&\boxcolorb{0.983}&\boxcolorb{0.777}\tabularnewline
    \hline
    \multicolumn{2}{|l|}{\CFP}&\boxcolorb{0.849}& & & \tabularnewline
    \hline
    \multicolumn{2}{|l|}{$\text{MSY}_{\text{theo}}$}&\boxcolorb{0.820}& & & \tabularnewline    
    \hline
    \multicolumn{2}{|l|}{Unfished}&\boxcolorb{0.982}& & & \tabularnewline    
       
    \hline

\end{tabular}

\end{table}

Table~\ref{tab:results} displays for each type of plan the average
proportion of $\text{MSY}_{\text{theo}}$ achieved in the MSE, together
with the proportion of surviving species and changes in size
structure.  Typical standard errors are around $3\%$ for yield scores,
$2\%$ for survival rates, and $0.12$ for mean log size.  More detailed
statistical results, including error estimates and statistical
significance of differences, are documented in \OM~\ref{sec:deta-quant-outc}.  There we also report yields for a
second, statistically independent MSE based on model communities of
half the size used here.
Remarkably, differences in yield scores between the two MSE
are small, suggesting that our findings depend little on the size of
communities.

Table~\ref{tab:results} 
shows that
exploitation leading to $\text{MSY}_{\text{theo}}$ would come at the
cost of conserving only about $72\%$ of fish species and a change in
species size structure corresponding a drop of $52\%$
($=1-10^{1.65-1.97}$) in geometric mean maturation body mass.
Such impacts are plausible in view of a corresponding survival rate of
$51\%$ obtained by \cite{Matsuda06:_MaximalYields}.  Yields from
\CFP{} were only about half of $\text{MSY}_{\text{theo}}$ and resulted
in survival of $63\%$ of species and a decline in geometric mean size
by $45\%$.

Among the full range of plans considered, a number of general trends
can be observed:
\begin{enumerate}
\item Independent of objective, \STC{} gave higher long-term yields
  than \PTC{}.  This is expected from structural instability
  \citep[Chapter 24]{rossberg13:_food_webs_biodiv}, as explained in
  \OM~\ref{sec:effect-struct-refpoints}. Remarkably, these
  yield increases were paralleled by higher species survival rates.
\item For the \TY{} objective, conservatism did substantially increase
  yields with both \PTC{} and \STC{}.  The same holds, to a lesser
  extent, for the \NP{} objective.  As theoretically expected from
  structural instability (\OM~\ref{sec:effects-struct-inst-par}
  and~\ref{sec:impl-cons-reg}), this was not the case for the \NS{}
  objective.  Conservatism increased species survival without
  exception.
\item With standard conservatism, \SOC{} resulted in yields
  intermediate between \STC{} and \PTC{}, independent of objective.
\item Optimal conservatism led to improvements over standard
  conservatism by 8.3--16.9\% of $\text{MSY}_{\text{theo}}$.  The 
  pattern of improvements was similar to that for the difference
  between standard and no conservatism.
\item Independent of 
objective, yields from \PTC{} using standard
  conservatism improved only slightly over \CFP{}.  However, yields
  from \PTC{} 
  with
  optimal conservatism were notably higher,
  which 
  indicates
  potential for some improvements through
  advances in mathematical methods. Without conservatism, yield score,
  survival rate and impact on community size structure for \PTC{}
  with \NP{} objective were almost identical to \CFP{}.
\item Independent of strategy, total yields obtained under \NS{}
  objective were consistently lower than those under \NP{} objective.
\item Remarkably, total yields attained aiming for \NP{} objective
  tended to be similar to or higher than those attained aiming for 
   \TY{} objective.  The expected reversal occurred only with the
  conservative variants of \STC{}. With optimal conservatism, \STC{}
  yield came (on average) within 7\% of $\text{MSY}_{\text{theo}}$,
  while survival rate and impact on community size structure were
  almost identical as for $\text{MSY}_{\text{theo}}$.
\item Impacts on species survival and community size structure were
  consistently lowest with \NS{}, highest with \TY{} and intermediate
  with \NP{} objective.
\end{enumerate}

\section{Discussion}
\label{sec:discussion}

\begin{table}
  \caption{Illustration of the impact of conservation constraints
    on the scores of selected management plans. \newline}
  \label{tab:conserv}

\newcommand{\boxcolor}[1]{\pgfmathparse{(1-(#1-0.3))}\xglobal\definecolor{mC}{rgb}{1,\pgfmathresult,\pgfmathresult}\cellcolor{mC}\color{black}\pgfmathparse{100*#1}\pgfmathprintnumber[precision=1,fixed zerofill=true,assume math mode=true]{\pgfmathresult}}

\newcommand{\boxcolorg}[1]{\pgfmathparse{(1-1.4*(#1-0.4))}\xglobal\definecolor{mC}{rgb}{\pgfmathresult,1,\pgfmathresult}\cellcolor{mC}\color{black}\pgfmathparse{100*#1}\pgfmathprintnumber[precision=1,fixed zerofill=true,assume math mode=true]{\pgfmathresult}}

\newcommand{\boxcolorb}[1]{\pgfmathparse{(1-1.4*(#1-0.5))}\xglobal\definecolor{mC}{rgb}{\pgfmathresult,\pgfmathresult,1}\cellcolor{mC}\color{yellow}\pgfmathparse{2.0102*#1}\pgfmathprintnumber[precision=2,fixed zerofill=true,assume math mode=true]{\pgfmathresult}}

\centering
\sf
\renewcommand{\arraystretch}{1.2}

\begin{tabular}{|l|l|l|l|l|}
\hline
\multirow{2}{*}{{\textbf{\begin{tabular}[c]{@{}l@{}}Scoring\\
        criterion\end{tabular}}}} &
\multicolumn{1}{c|}{\multirow{2}{*}{\begin{tabular}[c]{@{}c@{}}Management\\
      plan\end{tabular}}} &
\multicolumn{2}{c|}{\begin{tabular}[c]{@{}c@{}}Species\\
    conservation \end{tabular}} & \multirow{2}{*}{
  \begin{minipage}[b]{\widthof{$p$-values}}
    \centering
    paired\\ t-test\\ $p$-values
  \end{minipage}
} \\ 
\cline{3-4} & \multicolumn{1}{c|}{} & No & Yes&\\ 
\hline
\multirow{2}{*}{{\textbf{Yield}}} & \STC{} for \NP{} objective& \boxcolor{0.733} & \boxcolor{0.726} & 0.744 \\ 
\cline{2-4}  & \STC{} for \TY{} objective& \boxcolor{0.850} & \boxcolor{0.855} & 0.713 \\ 
\hline
\multirow{2}{*}{{\textbf{\begin{tabular}[c]{@{}l@{}}Survival\\ rate\end{tabular}}}} & \STC{} for \NP{} objective&\boxcolorg{0.776} &\boxcolorg{0.870} & 0\\ 
\cline{2-4}  & \STC{} for \TY{} objective& \boxcolorg{0.704} &\boxcolorg{0.919} & 0\\ 
\hline
\multirow{2}{*}{{\textbf{\begin{tabular}[c]{@{}l@{}}Mean\\ $\log_{10}$ size\end{tabular}}}} & \STC{} for \NP{} objective&\boxcolorb{0.879} &\boxcolorb{0.919} & 0.038\\ 
\cline{2-4}  & \STC{} for \TY{} objective&\boxcolorb{0.850} &\boxcolorb{0.945} & 0.004 \\ 
\hline

\end{tabular}

\end{table}

\subsection{Biodiversity impacts}
\label{sec:biodiversity-impacts}

In our MSE, strategies that yield more are also beneficial to species
survival and community size structure, while for management objectives
this relation is reversed.  While noteworthy, we caution not to
over-interpret these relations, because the management plans could be
modified to explicitly incorporate conservation constraints
\citep{Matsuda06:_MaximalYields}.  To illustrate possible implications
of conservation constraints, Table~\ref{tab:conserv} compares yield,
survival rate and community size structure with and without explicit
inclusion of such constraints in \STC{} management plans for \NP{} and
\TY{} objectives.  The constraint in this case is to not purposely
deplete fish populations to less than $10\%$ of their virgin biomass,
see \OM~\ref{sec:conservation}.  Interestingly, this modification
reduces extirpations and erosion of community size structure
substantially, without significantly affecting yields.  In view of this
robustness of yields, we disregard conservation in the following,
acknowledging that conservation while aiming at multispecies MSY
requires further study.


\subsection{Structural Instability}
\label{sec:struct-inst}

Our MSE demonstrates how structural instability of ecosystems, and so
of ecosystem models, can strongly impact management outcomes.  Results
from our MSE based on a complex food-web model closely follow
expectations derived from general mathematical arguments
\citep[Chapter~18; Supplementary
Material]{rossberg13:_food_webs_biodiv}.  The mathematical arguments
also appear to be robust to many real-world complications not
considered in the MSE, as already demonstrated for conservation
constraints.  Structural instability issues persist when transitioning
from stock- to fleet-based management
\citep{Rossberg12:_AttainingMMSY} or when multiplying yields by market
values.  Recruitment variability and imperfect control of applied
fishing pressures lead to additional uncertainty, which structural
instability can enhance (\OM~\ref{sec:effect-struct-refpoints}).

The MSE also highlighted options to mitigate these impacts.  The most
important may be adoption of state- rather than pressure targets in
management plans.  Another mitigation option is conservatism, i.e.\
regularization of the inverse problems underlying the computation of
multispecies MSY reference points.  The mathematical scheme for this
proposed in \OM~\ref{sec:impl-cons-reg} can be extended to more
detailed management models if required.  To determine good degrees of
conservatism when direct comparison of long-term management outcomes
is not practical, one can compare simulated outcomes using
members of a plausible model ensemble
\citep{Ianelli15:_MultimodelInference}, applying conservative reference
points derived with another model.

The EU's current approach to implementing its Common Fisheries Policy
(CFP, \citealt{eu13:_cfp}), here modelled as \CFP, is mathematically
similar to aiming at the \NP{} objective using \PTC{} without
systematic conservatism (\OM~\ref{sec:cfp-parameters}).  Our analysis
shows that this type of management plan is particularly vulnerable to
structural instability.  Expected yields are therefore comparatively
low.  These difficulties are illustrated in simulations by
\cite{lynam14:_expos_north_sea} predicting that a transition to the
MSY exploitation rates recommended by the International Council for
the Exploration of the Sea (ICES) will lower rather than raise total
long-term yields.

\subsection{Choice of objective}
\label{sec:choice-objective}

The MSE showed that the consequences of the two major policy choices
we considered, i.e.\ of objective and of strategy to achieve it, are
qualitatively independent.  They are therefore discussed separately
hereafter.  In doing so, we shall assume that management models do
take structural instability sufficiently into account such that the
``standard conservatism'' case in Tab.~\ref{tab:results} is
representative.


Among the three objectives analysed, sensitivity to structural
instability increases with increasing expected total yield, i.e.\ in
the expected order \NS{} $\to$ \NP{} $\to$ \TY{} objective
(theoretical expectations are summarised in
\OM~\ref{sec:summ-theor-expect}).  In view of the high empirical
uncertainty in the strengths of multispecies interactions, the \NS{}
objective might therefore be favoured by stakeholders, despite the
comparatively low yields this implies.  When combined with \STC{}, it
has the added advantages of transparency and simplicity; stakeholders
can focus on simply reaching agreement on the targeted ecosystem
state.  Exploitation rates and corresponding yields required to
achieve this state would play a role in negotiations, but not
explicitly become part of agreements.  Our results suggest this
approach would yield about 20-30\% higher than \CFP{}.

Despite \NP{} and \NS{} objective having similar definitions,
corresponding expected long-term yields clearly differ.  The higher
yields predicted for \NP{} could be reason for stakeholders to favour
this objective over \NS{}.  We therefore caution that we could not
find a reason why \NP{} should generally yield more.  Indeed, \OM~\ref{sec:summ-theor-expect} constructs counter-examples.  This
pattern should therefore be verified using MSEs that vary our setup.

Independent of the higher expected yields, there are two
considerations that favour \NP{} over \NS{} objective.  Firstly, \NP{}
could be more acceptable to stakeholders familiar with \CFP{} because
it is conceptually
similar.  Secondly, the
fictitious game 
underlying
the \NP{}
objective comes closer to 
the situation of real multi-stakeholder
fisheries than that underlying \NS{}.  Agreeing on this objective or
variations of it might therefore be easier.

Improvements in yields when going over from \TY{} to \NP{} objective
were surprisingly small, if present at all.  Even with \STC{}, they
were just 16\% (standard conservatism). This might be insufficient
reason to adopt the \TY{} objective, considering that this might lead
to difficult negotiations, as explained in the Introduction.
Nevertheless we anticipate that, when either \NP{} or \NS{} objective
is taken as the starting point for negotiations, they will develop
towards the \TY{} objective by agreements on modifications of targets
that give higher overall yields or incomes.

\subsection{Choice of strategy}
\label{sec:choice-strategy}

Considering the choice of strategy, out MSE strongly favours \STC{}
over \PTC{}.  This result is expected (\OM~\ref{sec:summ-theor-expect}), because structural instability
leads to large changes in community structure in response to small
changes (or errors) in applied pressures.  It is worth noting that,
for the same reasons, community structure is expected to be highly
sensitive to changes in the physical environment.  Within limits,
\STC{} counteracts this sensitivity.  We warn that MSE that do not use
a structurally unstable operating model, e.g.\ an operating model with
non-interacting species, will be unable to reproduce these phenomena!

While \PTC{} is currently favoured by the EU for most stocks, \STC{}
is either legal or \textit{de-facto} policy, e.g., in the U.S.A.,
Canada, and New Zealand.
Even the EU does not appear to be legally bound to \PTC{}.  The CFP in
its current form requires, according to Article~2, that ``management
[...]  restores and maintains populations of harvested species above
levels which can produce the maximum sustainable yield.'' It does then
set deadlines for adjusting exploitation rates accordingly, but these
are not the principal objective.  Further, the CFP requires
formulation of multiannual management plans with ``objectives that are
consistent with the objectives set out in Article~2 [...]''.  The
plans shall include ``quantifiable targets such as fishing mortality
rates and/or spawning stock biomass''.  However, considering the low
expected yields from \PTC{}, we suggest that multiannual plans setting
targets for fishing mortality or exploitation rates may not actually
be consistent with the objectives set out in Article~2.

\SOC{} strategies have their own strengths.  They are largely
insensitive to structural instability, do not require conservatism,
and yet achieve long-term yields nearly as large as those from
\STC{}. Further, the corresponding mHCRs are independent of
productivity rates ($r_i$ in Eq.~\eqref{eq:surplus-production}), which
lets these strategies automatically adjusts stock sizes to corrected
MSY levels if productivities change because of short-term or long-term
environmental change.

A specific disadvantage of \SOC{} with \TY{} objective, Transposed
Interactions Control, is that the prescribed exploitation rate of a
stock depends on the abundances of all stocks on which it has a direct
ecological effect: prey, predators, and competitors.  This can
increase measurement and planning uncertain. (For \STC{} such
dependencies enter through neutralising exploitation rates, but there
only as corrections to past observations, so uncertainty is lower.)
Despite its elegance, Transposed Interactions Control might therefore
be preferred only in exceptional cases, e.g.\ for data-rich
single-stakeholder fisheries that have already approached the \TY{}
objective and require a scheme to effectively adapt to uncertain
environmental change.

\subsection{Implementation}
\label{sec:implementation}

Most management plans we discussed require for their implementation
two models with different skills \citep{dickey-collas14:_hazar}.  The
first is what we called the management model.  It serves to compute
parameters of mHCRs consistent with objectives.  Emphasis is on the
skill to make long-term  
MSY projections.

In practice, a second type of model is needed to determine the stock
sizes $B_i$ required to compute exploitation rates $F_i$ from mHCRs
and to convert these into catch allowances.  Conventional age or
size-structured stock-assessment models can be use for this.  They
have the desired skill to accurately determine current states of
stocks based on current and historical data.  \STC{} requires in
addition that these models can estimate neutralising exploitation
rates.  To achieve this, they must be capable of forecasting states
for given exploitation rates a few years into the future.  The key
skill is therefore short-term projection.

Strategies to implement these two model types will be different.  For
long-term projections, interactions with the ecosystem at large and
general principles such as conservation and dissipation of energy as
it flows through food chains will be important.  For short-term
projections, optimal use of available data will be key, as with
current stock-assessment models.

\section{Conclusions}
\label{sec:conclusions}

Because in fact fish stocks do interact, it is not obvious how the
frequently invoked MSY policy objective should be translated to
real-world fisheries.  Answering questions~Q1 raised in Introduction,
we listed a range of management objectives that would for an isolated
stock all reduce to classical MSY.  Policy needs to clarify which one
is meant!  Addressing Q2, we considered a range of harvest control
rules and implementation details, and found that management outcomes
depend sensitively on their choice.

Answering Q3, the preceding Discussion highlighted trade-offs among
management options: the plans expected to generate highest yields
require large changes in management practice, accurate data, and/or
good collaboration among stakeholders.  The EU's current scheme of
\CFP{} yields substantially less than most other options
(Table~\ref{tab:results}), which answers~Q4.

While there are tradeoffs among all options considered, we propose as
a conceivable step towards increased effectiveness, sophistication,
and predictability of EU fisheries management a transition to \STC{}
with a \NP{} objective.  In the present study, we evaluated the scheme
$F= F_0 + (1-B_{\text{MSY}}/B)/T$, where $F_0$ is the neutralising
exploitation rate, $B$ is the current stock size, and $T$ is a
relaxation time parameter, arbitrarily chosen as $T=$1yr.  If the
resulting value for $F$ is negative, it is replaced by zero.  When
applied to an isolated species, the graph of $F$ vs.\ $B$ for the
corresponding HCR is similar, e.g., to that of the Fishery
Decision-Making Framework used by the Canadian government
(\url{http://www.dfo-mpo.gc.ca/fm-gp/peches-fisheries/fish-ren-peche/sff-cpd/precaution-eng.htm}),
except that $F$ is allowed to exceed the predicted value for
$F_{\text{MSY}}$.  Increases of $F$ beyond $F_{\text{MSY}}$ can be
necessary because $F_{\text{MSY}}$ is uncertain and an uncontrolled
stock size larger than the target value $B_{\text{MSY}}$ could lead,
through multispecies interactions, to unintended and uncontrolled
effects on other stocks, a situation that \STC{} aims to avoid.

Variations of this mHCR are conceivable, and many may be similarly
effective.  For example, following discussion within the Working Group
on Multispecies Assessment Methods of ICES \citep{ices14:_wgsam}, we
evaluated the modified rule
$F= F_0 - |F_{\text{SSC}}|(1-B/B_{\text{MSY}})$, where
$F_{\text{SSC}}$ is the target exploitation rate of the \CFP{} scheme.
While with this modified rule target stock size is approached at the
potentially slower rate $|F_{\text{SSC}}|$, the rule has the
intriguing property that it would reduce to $F=F_{\text{SSC}}$, i.e.\
to conventional \CFP{}, if the single-species Schaefer model was
accurate.  Reality is more complicated and the HCR will deviate from
\CFP{}, but these deviations may be small.  Applying our MSE to this
rule with a \NP{} objective, we obtain on average long-term total
yield of 75.4\% $\text{MSY}_{\text{theo}}$ (standard conservatism),
not significantly different from the result using our original \STC{}
rule ($p=0.15$, paired t-test).  Such results open up the possibility
of swiftly transitioning from the current scheme to \STC{}, enabling
higher future yields and more predictable ecological outcomes, with
minimal changes to currently recommended exploitation rates.


As discussed above, the formulation of EU's CFP is open to both \PTC{}
and \STC{}.  We recommend using of this flexibility to achieve best
possible outcomes for society and the marine environment.

\begin{notes}[Supplementary material]
  As Supplementary material, we provide a PDF file with mathematical
  definitions and analyses of management objectives and strategies,
  details of the methods used to implement these in our MSE, and
  statistical details of results.
\end{notes}

\begin{notes}[Acknowledgements]
  We thank Georgia Bayliss-Brown for comments on our manuscript.  This
  work was supported by the European Community's Seventh Framework
  Programme (FP7/2007-2013) under grant agreements no. 289257
  (MYFISH), the UK Department of Environment, Food and Rural Affairs
  (M1228), and the Natural Environment Research Council and the UK
  Department for Food, Environment and Rural Affairs within the Marine
  Ecosystems Research Program (MERP, NE/L00299X/1).
\end{notes}

\bigskip


\putbib{}

\end{bibunit}

\clearpage
\newpage

\renewcommand{\thefigure}{S\arabic{figure}}
\renewcommand{\thetable}{S\arabic{table}}
\renewcommand{\theequation}{S\arabic{equation}}
\renewcommand{\thesection}{S\arabic{section}}
\renewcommand{\thesubsection}{S\arabic{section}.\arabic{subsection}}
\setcounter{equation}{0}%
\setcounter{section}{0}%
\setcounter{subsection}{0}%

\begin{center}
  \vspace*{\fill}
  \large \textbf{Supplementary Material}
  \vspace*{\fill}
  \vspace*{\fill}
\end{center}
\begin{bibunit}

  These supplementary materials mainly provide two kinds of
  information.  (A) Mathematical formulations and analyses of
  management objectives and strategies and (B) further details and
  formal descriptions of the MSE reported in the paper.
  Sections~\ref{sec:msy-pdmm} to~\ref{sec:summ-theor-expect}
  incrementally build upon each other.  To follow the arguments it is
  therefore advised to read the material sequentially.
  Section~\ref{sec:deta-quant-outc} provides additional technical
  information on the MSE, and is largely independent of the previous
  sections.

\section{Management objectives}
\label{sec:msy-pdmm}

We provide formal mathematical descriptions of the three major types
of objectives considered in the main text, \NP{}, \NS{}, and \TY{}.
Among the possible ways to achieve these objectives, only idealised
community steady states with fixed stock biomasses $B_i^*$
($i=1,\ldots,S_F$) shall here be considered.  Even when such states
are not stable for a given and rigidly fixed set of exploitation rates
$F_i$, it will often be possible to stabilise it by letting
exploitation rates vary sightly around their long-term averages
according to appropriate harvest control rules.  Thus, oscillatory
states, which are economically undesirable, can be avoided and need
not be considered here.

With the long-term yield per unit time interval from species $i$ given
by $Y_i=B_i^*F_i$ ($i=1,...,S_F$), the corresponding total yield is
$Y=\sum_i^{S_F} Y_i$.  To keep the formalism simple, we assume that
the relationship between combinations of (averaged) exploitation rates
$F_i$ and attainable steady states $B_i^*$ is one-to-one, so that
community steady states and the corresponding yields can be
parameterized in terms of the corresponding exploitation rates $F_i$.
Coverage of ambiguities in the description of states by corresponding
pressures would inflate the mathematical formalism without adding much
understanding of use for the present work.

Mathematically, the \TY{} objective is attained with a set of
exploitation rates $F_i=F_{\text{MSY},i}$ such that
\begin{align}
  \label{eq:ty0}
  Y(F_{\text{MSY},1},...,F_{\text{MSY},S}) &=
  \max_{\mathbf{F}\in\mathbb{R}^{S_F}_{\ge0}} Y(F_1,...,F_S)\qquad\text{(\TY)}.
\end{align}
As conventional, $\mathbb{R}_{\ge0}$ denotes the set of all
non-negative real numbers and $\mathbb{R}^{S_F}_{\ge0}$ a set of
$S_F$-component vectors of non-negative real numbers.  Specifically,
$\mathbf{F}$ denotes the vector of fishing mortalities $F_i$.  

The \NP{} objective is attained for a set of fishing mortality rates
$F_i=F_{\text{MSY},i}\in \mathbb{R}^{S_F}_{\ge0}$ such that
\begin{align}
  \label{eq:np0}
  Y_i(F_{\text{MSY},1},...,F_{\text{MSY},S}) &=
  \max_{F_i\in\mathbb{R}_{\ge0}}
  Y_i(F_{\text{MSY},1},...,F_i,...,F_{\text{MSY},S})\qquad\text{(\NP)}
\end{align}
for all $1 \le i \le S$.  The \NS{} objective is best defined
using a parameterization in terms of corresponding states
$B_i^*=B_{\text{MSY},i}\ge 0$: for all $1 \le i \le S$,
\begin{align}
  \label{eq:ns0}
    Y_i(B_{\text{MSY},1},...,B_{\text{MSY},S}) &= \max_{B_i^*\in\mathbb{R}_{\ge0}}
  Y_i(B_{\text{MSY},1},...,B_i^*,...,B_{\text{MSY},S})\qquad\text{(\NS)}.
\end{align}

\section{The Operating Model}
\label{sec:popul-dynam-match}

\subsection{Model structure and parameterization}
\label{sec:model-struct-param}

The Population-Dynamical Matching Model (PDMM) generates large model
communities using an iterative process representing the natural
assembly of complex ecological communities. A full description of the
model equations, including all parameters and variables, is given in
\citeauthor{rossberg13:_food_webs_biodiv}
(\citeyear{rossberg13:_food_webs_biodiv}, Chapter 22). We describe
here only the model's main features.

A PDMM model community consists of $S=S_P+S_C$ species of which $S_P$
are producers and $S_C$ are consumers. Each species $j$ is
characterised by its adult body mass and a 5-component vector of
abstract vulnerability traits, and consumers in addition by a
5-component vector of abstract foraging traits. The trait vectors
locate species in trophic niche space
\citep{nagelkerke14:_niche_image,
  rossberg09:_how_troph_inter_stren_depen_trait}. They are thought to
be related in an undescribed way to the phenotype of species.  The
state of the population of each species $j$ is characterised by its
population biomass $B_j$ alone. 

For any conceivable consumer-resource pair, their trophic interaction
strength depends on their body mass ratio and the relative position of
their traits in trophic niche space.  Interaction strengths are large
if the producer is in the consumer's preferred size range and if the
vulnerability trait vector of the producer is close to the position of
the consumer's foraging traits vector. The available trophic niche
space is restricted by constraining the maximum length of
vulnerability traits vectors.

Starting from a few producer and consumer species, larger communities
are assembled by alternately adding new species that differ from a
resident species by a small change in adult body mass and traits, and
then simulating the population dynamics of the system to a new
equilibrium. The species that go extinct in the extended community are
removed in each iteration.

Community dynamics is formulated through a set of coupled ordinary
differential equations, each representing the dynamics of one
species' population. In the most generic form, this system of ODEs can
be expressed by Eq.~\eqref{eq:pdmm-general} of the main text.

In system~\eqref{eq:pdmm-general}, the linear growth rates
$g_i(B_1,\dots,B_S)$ are non-linear functions of their arguments,
including extended type II functional responses, which are nonlinear
functions of the biomasses $B_1,\linebreak[1]\dots,\linebreak[1]B_S$.
Details are as in \citeauthor{rossberg13:_food_webs_biodiv}
(\citeyear{rossberg13:_food_webs_biodiv}, Chapter 22), with
modifications as given at the end of this section. These type II
functional responses complicates the interpretation of model outputs,
compared to simpler interaction terms, but have been shown to lead to
dynamically stable food-webs with topologies similar to those seen in
natural terrestrial and aquatic communities, which are otherwise
difficult to build.

As shown by \cite{rossberg13:_food_webs_biodiv}, the size of PDMM
communities scales approximately linearly with a parameter $X$ that
controls both the probability of direct competition among producers
and the maximum length of vulnerability traits vectors. The outcome
of, say, doubling $X$ is to halve the probability of direct
competition among producers and to double the volume of trophic niche
space, which is similar to simply modelling two communities in
parallel and then ``rewiring'' interactions to go criss-cross between
the two communities.  Results reported in Table~\ref{tab:results} of
the main text correspond to the choice $X=2$, for comparison we also report
corresponding results for $X=1$.

For $X=2$ and with the model definition used by
\cite{rossberg13:_food_webs_biodiv} we found that producers tended to
evolve increasingly larger body sizes, which is undesirable for
studies of open-ocean communities.  We avoided this by going back to
a model formulation of \cite{rossberg07:_flat}, where producer
population growth rates are given by
\begin{align}
  \label{eq:exp-growth}
  g_i(B_1,\dots,B_S)=s_i
\frac{ \displaystyle \left[
    \exp
    \left(
      -\sum_{\text{producers j}} d_{ij} B_j
    \right) - l
  \right]}{1-l} - \text{predation},
\end{align}
with $l=0.5$, rather than using the Lotka-Volterra-type model
\begin{align}
  \label{eq:lin-growth}
  g_i(B_1,\dots,B_S)=s_i
    \left[
    1-\sum_{\text{producers j}} d_{ij} B_j
  \right] - \text{predation}
\end{align}
of \cite{rossberg13:_food_webs_biodiv}.  To compensate for the
resulting slightly lower primary production, we further had to scale
up all consumer attack rates by a factor $5/4$---see
\cite{rossberg12:_analytic_size_spectrum,rossberg13:_food_webs_biodiv}
for the rationale underlying such fine-tuning of attack rates.

  
  
  


\subsection{The evolutionary search algorithm used to maximise total
  yield}
\label{sec:modif-evol-search}

Next we describe the method used to find the theoretically maximal
total yield $\text{MSY}_{\text{theo}}$ for each model community.  For
the two types of Nash objectives, no attempts were made to find the
corresponding solutions in the operating model.  Numerical methods to
find Nash equilibria seem to be much less developed than algorithms to
maximise a single goal function, such as total equilibrium yield $Y$.
Even the latter problem is formidable in our case.  From experience
testing different approaches, it appears that the search space might
best be visualised to have the structure of a mountain landscape such
as the Alps.  The main maximum is surrounded by several side maxima of
similar magnitude, and such local maxima may not be connected to each
other by steep ridges. Further off, lower side-maxima can be found.
Such goal functions make numerical optimisation difficult.  Standard
hill-climbing algorithms, for example, almost inevitably get caught in
low local maxima corresponding to the Prealps.

To find the fishing mortalities achieving the \TY{} objective,
Eq.~\eqref{eq:ty0}, we therefore used the Improved Stochastic Ranking
Evolution Strategy (ISRES) of \cite{runarsson05:_searc}, as
implemented in \texttt{nlopt} v2.3
(\url{http://ab-initio.mit.edu/wiki/index.php/NLopt}).  The
optimisation library \texttt{nlopt} provides implementations of
several optimisation algorithms, easily accessible through various
programming languages.  We describe ISRES here in some detail, because,
at one crucial point, we had to modify it.

ISRES is an evolutionary search algorithm that iteratively computes
the goal function for a population of parameters combinations $x_i$
($1\le i\le n$, here $x_i=\ln F_i$ and $n=S_F$), chooses a sub-set of
``surviving'' good combinations, and then obtains the next generation
of the population by randomly mutating the parameters of survivors
(there is no recombination step).  Together with the parameters $x_i$
themselves, the mutation step sizes are inherited and mutated along
lineages.  Specifically, parameter $x_i'$ and its mutation step sizes
$\sigma_i'$ in the next generation are computed from the values $x_i$
and $\sigma_i$ of a survivor of the current generation according to
the rule
\begin{subequations}
  \begin{align}
    \label{eq:sigpp}
    \hat\sigma_i&=\sigma_i \exp
    \left(
      \tau'\xi+\tau \xi_i
    \right),\\
    x'_i&=x_i+\hat\sigma_i \xi'_i,\\
    \label{eq:smoothing}
    \sigma'_i&=\sigma_i+\alpha\cdot(\hat\sigma_i-\sigma_i),
  \end{align}
\end{subequations}
where $\xi$, $\xi_i$, and $\xi_i'$ ($1\le i \le n$) are independent
random numbers with standard-normal distributions, and the parameters
$\tau$ and $\tau'$ are chosen following
\cite{schwefel93:_evolut_optim_seekin}.  The last step,
Eq.~\eqref{eq:smoothing}, with $\alpha=0.2$, follows a proposal by
\cite{runarsson02:_reduc}.  It controls the degree of heritability of
the ``phenotype'' $\hat \sigma_i$ of mutation step size along
lineages, and so provides for some memory and accumulation of
experience with different mutation step sizes as lineages evolve.

The problem we encountered with this scheme is that
Eq.~\eqref{eq:smoothing} biases mutations of $\ln\sigma_i$ towards
larger values in the absence of selection pressures.  It so undermines
the intention of Eq.~\eqref{eq:sigpp} to find the appropriate order of
magnitude of $\sigma_i$ by conducting a free search on the
$\ln\sigma_i$ axis until selection sets in because a good value of
$\sigma_i$ is found.  Starting with large values of $\sigma_i$, the
rugged goal function of our problem does not provide sufficient cues
for good mutation step sizes to counteract this bias, so that step
sizes remain large and the algorithm never converges.

We overcame this problem by replacing Eq.~\eqref{eq:smoothing} with a
corresponding exponentially declining memory on the
$\log\sigma_i$-scale.  Specifically, we set
\begin{align}
  \label{eq:sigppx}
  \sigma'_i&=\sqrt{\hat\sigma_i \sigma_i}. \tag{\ref{eq:smoothing}'}
\end{align}

For each iteration of the search algorithm, total yield $Y$ was
computed from the steady state reached after starting simulations from
the virgin community and fishing it at constant exploitation rates
$F_i$.  To simplify calculations, no attempt was made to stabilise
dynamics when oscillatory attractors were reached.  Rather, we
averaged yield $Y$ in this case over a sufficiently long time interval
to obtain the mean long-term yield.

\section{Calibrating the management model}
\label{sec:calibr-mang-model}

The idea behind our calibration of the management model, i.e., of the
multispecies Schaefer model in Eq.~\eqref{eq:surplus-production} of
the main text, is to perform first an adiabatic elimination of all
non-fish species from the PDMM, and then to choose parameters of the
Schaefer model such that it reproduces the dynamics of the PDMM to
linear order in the distance from the PDMM's current state.

Technically, this is achieved to a good approximation by first
numerically computing the interaction matrix $\mathbf{G}'$
\citep{berlow04:_interac_strength} of the PDMM model community, given
by its entries
\begin{align}
  \label{eq:Gprime}
  G_{ij}'=-\frac{\partial g_i(B_1,\ldots,B_S)}{\partial B_j},
\end{align}
and then dividing it into blocks
\begin{align}
  \label{eq:Gprime-blocks}
  \mathbf{G}'=
  \left(
    \begin{matrix}
      \mathbf{G}_{\text{FF}}&\mathbf{G}_{\text{FN}}\\
      \mathbf{G}_{\text{NF}}&\mathbf{G}_{\text{NN}}
    \end{matrix}
  \right)
\end{align}
pertaining to direct interactions within and between the communities
of fish (F) and non-fish (N) species.  Elimination of the non-fish
species is achieved by computing the effective $S_F\times S_F$
interaction matrix for fish
\begin{align}
  \label{eq:G}
  \mathbf{G}=\mathbf{G}_{\text{FF}}-\mathbf{G}_{\text{FN}}\mathbf{G}_{\text{NN}}^{\smash{-1}}\mathbf{G}_{\text{NF}}.
\end{align}
The linear growth rates $r_i$ entering the Schaefer model are then
obtained as
\begin{align}
  \label{eq:ri}
  r_i=g_i(B_1,\ldots,B_S)+\sum_j^{S_F} G_{ij}B_j+F_i.
\end{align}
The resulting Schaefer model can be written in matrix notation as 
\begin{equation}
  \label{LVmat} \frac{d\bf B}{dt}=\left[{\bf r}-{\bf GB}-{\bf F}\right]\circ
  {\bf B}
\end{equation} where $\circ$ denotes the Hadamard or the entrywise
product.

\section{Effects of structural instability of the 
  operating model on
  the management model}
\label{sec:effects-struct-inst-mod}

We now sketch some considerations relating structural instability of
the operating model to properties of the management model.
Mathematically versed readers will note that we gloss over several
technical intricacies which could presumably be ironed out through
more elaborate technical arguments.

For structurally unstable ecological models, the ecological
interaction matrix tends to have several eigenvalues $\lambda$ near
zero \citep{rossberg13:_food_webs_biodiv}.  Hence, there are
$S$-component vectors $\mathbf{e}$ of lengths $|\mathbf{e}|=1$ such
that
\begin{align}
  \label{eq:small-lambda}
  \mathbf{G}'\mathbf{e}=\lambda \mathbf{e}\approx 0
\end{align}
because $\lambda \approx 0$.  This phenomenon persists even after
scaling out differences in the characteristic populations sizes and
rates of change for different ecological groups, as achieved by going
over to a representation in terms of the competitive overlap matrix
\citep{rossberg13:_food_webs_biodiv}, which is why we shall disregard
such differences for the present considerations.  Dividing
$\mathbf{G}'$ into blocks as in Eq.~\eqref{eq:Gprime-blocks} and
correspondingly writing
\begin{align}
  \label{eq:v-blocks}
  \mathbf{e}=
  \left(
    \begin{matrix}
      \mathbf{e}_{\text{F}} \\ \mathbf{e}_{\text{N}}
    \end{matrix}
  \right),
\end{align}
Eq.~\eqref{eq:small-lambda} becomes
\begin{subequations}
    \label{eq:small-lambda-blocks}
    \begin{align}
      \label{eq:slb-F}
      \mathbf{G}_{\text{FF}}\mathbf{e}_{\text{F}}+\mathbf{G}_{\text{FN}}\mathbf{e}_{\text{N}}\approx 0,\\
      \label{eq:slb-N}
      \mathbf{G}_{\text{NF}}\mathbf{e}_{\text{F}}+\mathbf{G}_{\text{NN}}\mathbf{e}_{\text{N}}\approx
      0.
    \end{align}
\end{subequations}

Now one needs to distinguish two cases.  The first is given by
$|\mathbf{e}_{\text{F}}|^2\ll 1$ and
$|\mathbf{e}_{\text{N}}|^2\approx 1$, in the second
$|\mathbf{e}_{\text{F}}|^2$ is of the order of magnitude of one.  In
the first case, Eq.~\eqref{eq:slb-N} reduces to
$\mathbf{G}_{\text{NN}}\mathbf{e}_{\text{N}}\approx 0$, implying that
$\mathbf{G}_{\text{NN}}$ has eigenvalues near zero. Matrices that have
eigenvalues near zero and others further away are said to be
ill-conditioned.  The inverse of such a matrix,
e.g. $\mathbf{G}_{\text{NN}}^{-1}$, depends sensitively on the values
of matrix entries.  Since $\mathbf{G}_{\text{NN}}^{-1}$ enters the
effective interaction matrix for fish through Eq.~\eqref{eq:G}, this
can degrade the accuracy of $\mathbf{G}$ in describing interactions
among fish.

In the second case, elimination of $\mathbf{e}_{\text{N}}$ from
Eq.~\eqref{eq:slb-F} using Eq.~\eqref{eq:slb-N} leads to
\begin{align}
  \label{eq:eF-is-nullv}
  0\approx \mathbf{G}_{\text{FF}}\mathbf{e}_{\text{F}}-\mathbf{G}_{\text{FN}}\mathbf{G}_{\text{NN}}^{\smash{-1}}\mathbf{G}_{\text{NF}}\mathbf{e}_{\text{F}}=\mathbf{G}\mathbf{e}_{\text{F}},
\end{align}
implying that the effective interaction matrix $\mathbf{G}$ itself has
eigenvalues near zero.  As a result, the Schaefer model
Eq.~\eqref{LVmat} is then structurally unstable.

Thus, structural instability of the operating model can make the
management model \eqref{LVmat} an inaccurate description of the
interactions among fish or translate into structural instability of
the management model itself.  For large, complex operating models, as
in our case, one can expect both cases to play out to varying degrees.
With the more elaborate management models used in practice, this
inheritance of structural instability is likely to prevail in similar,
albeit less transparent, form.

\section{Solving the management model for MSY objectives and mHCR
  parameters}
\label{sec:solv-oper-model}

\subsection{Standard case}
\label{sec:stand-case-unconstr}

We now derive the parameters entering the mHCR in Section~2.3 of the main
text for the objectives that we consider.  Setting aside the
possibility of extirpations ($B_i^*=0$), equilibrium solutions of
Eq.~\eqref{LVmat} satisfy
\begin{equation}
  \label{equilF} {\bf F}={\bf r}-\mathbf{G}\mathbf{B}^*
\end{equation}
We assume that $\mathbf{G}$ is non-singular, even though it might be
ill-conditioned.  Equation~\eqref{equilF} then implies immediately the
relations
\begin{align}
  \label{eq:FB-partials}
  \frac{\partial F_i}{\partial B_j^*}=-G_{ij}\quad\text{and}\quad\frac{\partial B_i^*}{\partial F_j}=-G_{ij}^{-1},
\end{align}
which we shall use in the following.

Making use of the continuity of the derivatives in
Eq.~\eqref{eq:FB-partials}, one obtains, using standard arguments,
necessary conditions for satisfaction of the objectives described in
Section~\ref{sec:msy-pdmm}.  For the \TY{} objective these are, with
$1\le i \le S_F$,
\begin{align}
  \label{eq:ty-analytic}
  0=\frac{\partial Y}{\partial F_i}=\frac{\partial \sum_j F_j
    B_j^*}{\partial F_i}=B_i^*+\sum_jF_j\frac{\partial B_j^*}{\partial
    F_i},\quad\text{i.e.}\quad 0={\mathbf{B}^*}\tr-\mathbf{F}\tr\mathbf{G}^{-1},
\end{align}
(with $\tr$ denoting matrix transposition); for the \NP{} objective
\begin{align}
  \label{eq:np-analytic}
  0=\frac{\partial Y_i}{\partial F_i}=\frac{\partial F_i
    B_i^*}{\partial F_i}=B_i^*+F_i\frac{\partial B_i^*}{\partial
    F_i},\quad\text{i.e.}\quad 0={\mathbf{B}^*}\tr-\mathbf{F}\tr\diag(\mathbf{G}^{-1}),
\end{align}
with $\diag(\mathbf{M})$ denoting the diagonal matrix with diagonal
entries given by those of $\mathbf{M}$; and for the \NS{} objective
\begin{align}
  \label{eq:ns-analytic}
  0=\frac{\partial Y_i}{\partial B_i}=\frac{\partial F_i
    B_i^*}{\partial B_i^*}=\frac{\partial F_i}{\partial
    B_i^*}B_i^*+F_i,\quad\text{i.e.}\quad 0=-\diag(\mathbf{G})\mathbf{B}^*+\mathbf{F}.
\end{align}
All three conditions can be re-written as
\begin{align}
  \label{eq:G-hat-cond}
  \mathbf{F}=\hat{\mathbf{G}}\tr \mathbf{B}^*,
\end{align}
where for the \TY{} objective $\hat{\mathbf{G}}=\mathbf{G}$, for \NP{}
$\hat{\mathbf{G}}=\diag(\mathbf{G}^{-1})^{-1}$, and for \NS{}
$\hat{\mathbf{G}}=\diag(\mathbf{G})$.

The \SOC{} rules are obtained as
\begin{align}
  \label{eq:soc-general}
  \mathbf{F}=\hat{\mathbf{G}}\tr \mathbf{B},
\end{align}
by simply replacing the equilibrium values $\mathbf{B}^*$ with the
current values $\mathbf{B}$ of stock sizes in
Eq.~\eqref{eq:G-hat-cond}.  State targets
$\mathbf{B}^*=\mathbf{B}_{\text{MSY}}$ are obtained by putting
Eq.~\eqref{eq:G-hat-cond} into Eq.~\eqref{equilF} and solving for
$\mathbf{B}^*$:
\begin{align}
  \label{eq:BMSY}
  \mathbf{B}_{\text{MSY}}=
  \left(
    \mathbf{G}+\hat{\mathbf{G}}\tr
  \right)^{-1}\mathbf{r}.
\end{align} 
Pressure targets $\mathbf{F}=\mathbf{F}_{\text{MSY}}$ are obtained by
combining Eqs.~\eqref{eq:G-hat-cond} and \eqref{eq:BMSY}:
\begin{align}
  \label{eq:FMSY}
  \mathbf{F}_{\text{MSY}}= \hat{\mathbf{G}}\tr
  \mathbf{B}_{\text{MSY}}= \hat{\mathbf{G}}\tr\left(
    \mathbf{G}+\hat{\mathbf{G}}\tr \right)^{-1}\mathbf{r}.
\end{align} 

\subsection{Modifications to incorporate conservation constraints}
\label{sec:conservation}

Results in Table~\ref{tab:conserv} of the main text were computed using
variants of Eq.~\eqref{eq:BMSY} and Eq.~\eqref{eq:FMSY} that take
conservation constraints into account.  These are given by the
following algorithm.

For each species $i$, a conservation threshold biomasses
$B_{\text{cons},i}$ is first defined as $10\%$ of the species' biomass
in the unexploited community \citep{matsuda06:_maxim_yield}.  If
$B_{\text{MSY},i}$, as computed using Eq.~\eqref{eq:BMSY}, falls below
this threshold for one or more species, the species $i$ with the
lowest ratio $B_{\text{MSY},i}/B_{\text{cons},i}$ is identified and
its target abundance set to $B_{\text{cons},i}$.  Fixing
$B_i=B_{\text{cons},i}$ in Eq.~\eqref{LVmat} and eliminating species
$i$ from the system leads to a new Schaefer (or Lotka-Volterra) model
of the same form as Eq.~\eqref{LVmat} with one fewer species. For this
model Eq.~\eqref{eq:BMSY} is re-evaluated.  This procedure is repeated
until $B_{\text{MSY},i}>B_{\text{cons},i}$ for all remaining species.
Equation~\eqref{eq:FMSY} is then evaluated with $B_{\text{MSY},i}$
replaced by $B_{\text{cons},i}$ where applicable.

\section{Target fishing mortalities for \CFP{}}
\label{sec:cfp-parameters}

To derive a HCR for \CFP{} of the $i$th fish species, we proceeded
essentially along the same route as that leading to
Eq.~\eqref{eq:FMSY}, but now adiabatically eliminating all species but
species $i$ to obtain the single-species case of the Schaefer model
Eq.~\eqref{LVmat}.  To obtain the non-linear coefficient $\tilde{g}_i$
corresponding to the interaction matrix $\mathbf{G}$ in the
single-species model
\begin{align}
  \label{eq:logistic}
  \frac{dB_i}{dt}= \left[ \tilde r_i-\tilde g_iB_i-F_i \right]B_i,
\end{align}
observe that $\mathbf{G}^{-1}$, as defined through Eq.~\eqref{eq:G},
is the upper left $S_F\times S_F$ block of the inverse of
$\mathbf{G}'$ \citep{bernstein05:_matrix_mathem}.  Correspondingly,
$\tilde{g}^{-1}_i$ is the $1\times 1$ block on the diagonal of
${\mathbf{G}'}^{-1}$ corresponding to species $i$, which this is also
the $i$th diagonal element of $\mathbf{G}^{-1}$.  The coefficient
$\tilde r_i$ for the linear growth rate in Eq.~\eqref{eq:logistic} is
obtained following the idea of Eq.~\eqref{eq:ri}, for simplicity
assuming equilibrium ($g_i(...)=0$), as $\tilde r_i=\tilde g_i B_i+F_i$.  The
combined Schaefer models so obtained for the $S_F$ fish species can be
summarised in matrix notation as
\begin{align}
  \label{eq:logistics}
  \frac{d\mathbf{B}}{dt}=
  \left[
    \tilde{\mathbf{r}} - \tilde{\mathbf{G}}\mathbf{B} -\mathbf{F}
  \right]\circ \mathbf{B}
\end{align}
with $\tilde{\mathbf{G}}=\diag(\mathbf{G}^{-1})^{-1}$, which is
identical to $\hat{\mathbf{G}}$ for the \NP{} objective.

Managers assuming validity of these $S_F$ single-species models will
compute MSY target exploitation rates as
\begin{align}
  \label{eq:F-SSC}
  \mathbf{F}_{\text{SSC}}=\frac{\tilde{\mathbf{r}}}{2}.
\end{align}
Even though Eq.~\eqref{eq:logistics} disregards multispecies
interactions, adaptive management in which the parameters of
Eq.~\eqref{eq:logistics} are iteratively adjusted might eventually
converge to a community state $\mathbf{B}^*$ where indeed, with
$\mathbf{F}=\mathbf{F}_{\text{SSC}}$, the equilibrium condition of
Eq.~\eqref{eq:logistics},
\begin{align}
  \label{eq:0-SSC}
  \tilde{\mathbf{G}}\mathbf{B}^*=\tilde{\mathbf{r}}-\mathbf{F}=\mathbf{F},
\end{align}
is met.  The community then satisfies the \NP{} objective, as can be
verified by noting that Eq.~\eqref{eq:0-SSC} is logically equivalent
to Eq.~\eqref{eq:G-hat-cond} with
$\hat{\mathbf{G}}=\tilde{\mathbf{G}}=\tilde{\mathbf{G}}\tr$.  \CFP{}
for MSY can therefore be understood as a technically simple route to
attaining the \NP{} objective.

\section{Effects of structural instability of the management model on mHCR parameters}
\label{sec:effects-struct-inst-par}

How do structural instability of the management model and parameter
uncertainty in $\mathbf{G}$ (derived in
Sec.~\ref{sec:effects-struct-inst-mod}), affect the accuracy at which
the parameters of the mHCR, as derived in
Sec.~\ref{sec:solv-oper-model}, can be computed?  To answer this
question, we first have to develop further intuition about the
structure of the set of the $S_F$ eigenvalues of $\mathbf{G}$.
Because $\mathbf{G}$ is not generally symmetric, its eigenvalues are
not generally real numbers---they are complex numbers, represented by
points in the complex plane, given by two coordinates: their real and
their imaginary parts.  Because the entries of $\mathbf{G}$ are real
numbers, the distribution of its eigenvalues over the complex plane is
symmetric with respect to reflection on the real axis.

The eigenvalues of asymmetric random matrices tend to be distributed
more or less evenly over limited areas of the complex plane, and one
can expect this to be similarly the case for $\mathbf{G}$.  For
ecological model communities built through an assembly process, one
finds that this area is convex and comes close to or touches the
origin of the complex plane (which corresponds to the eigenvalue
zero), but does not fully cover it, thus marking the onset of
structural instability \citep{rossberg13:_food_webs_biodiv}. By the
sign convention for $\mathbf{G}$ used here, this area is located in
the right complex half-plane, where the real parts of eigenvalues are
positive.

The formula for $\mathbf{B}_{\text{MSY}}$, Eq.~\eqref{eq:BMSY}, can be
rewritten as $\mathbf{B}_{\text{MSY}}=\mathbf{H}^{-1}\mathbf{r}/2$
with
\begin{align}
  \label{eq:H-def}
  \mathbf{H}=\frac{\mathbf{G}+\hat{\mathbf{G}}\mathrlap{\tr}}{2}.
\end{align}
The condition of $\mathbf{H}$, and so the impact of inaccuracies in
its entries on the accuracy of the reference points
$\mathbf{B}_{\text{MSY}}$ and $\mathbf{F}_{\text{MSY}}$ (\textit{via}
Eq.~\eqref{eq:FMSY}), depends on $\mathbf{G}$ and on the definition of
$\hat{\mathbf{G}}$ in terms of $\mathbf{G}$ pertinent to the
particular objective.  The decisive question is how the operation on
$\mathbf{G}$ defined by Eq.~\eqref{eq:H-def} transforms the area
covered by its eigenvalues in the complex plane. This depends on the
particular management objective chosen.  One can expect that the
condition of $\mathbf{H}$ becomes poor if the transformed area covers
the origin, i.e. if $\mathbf{H}$ has eigenvalues with both negative
and positive real parts, and that its condition becomes good if the
transformed area gets detached from the origin.

A first point to note is that, because the sum of all eigenvalues of a
matrix equals the sum of its diagonal elements, the respective
arithmetic means are equal as well.  The ``centre of gravity'' of the
eigenvalues of $\mathbf{H}$ in the complex plane is therefore
identical to that of $\mathbf{G}$ when the diagonal entries of the two
matrices are identical, as is the case for \TY{} and \NS{} objective
(where $\hat{\mathbf{G}}=\mathbf{G}$ and
$\hat{\mathbf{G}}=\diag(\mathbf{G})$, respectively).

In case of the \TY{} objective ($\hat{\mathbf{G}}=\mathbf{G}$),
$\mathbf{H}$ is simply the symmetric part of $\mathbf{G}$.  The
eigenvalues of the symmetric part of a matrix are always spread out at
least as widely along the real axis as those of the matrix itself
(see, e.g., \citealt{bhatia97:_matrix_analy}, pp. 73-74, presented
more accessibly by \citealt{raghavan06:_stabil_system_random_param}),
and generally wider. Because typically the set of eigenvalues of
$\mathbf{G}$ comes close to the origin, one can so expect that the
eigenvalues of $\mathbf{H}$ cover the origin, implying that
$\mathbf{H}$ is ill-conditioned.

For the \NS{} objective ($\hat{\mathbf{G}}=\diag(\mathbf{G})$),
Eq.~\eqref{eq:H-def} amounts to reducing all off-diagonal elements of
$\mathbf{G}$ to half of their values.  The effect of this is best
understood by first considering the impact on eigenvalues of the
opposite transformation: gradually adding off-diagonal elements to a
diagonal matrix.  A simple perturbation analysis
\citep[e.g.,][]{bloch12:_level_qcd}, strictly valid only when the off-diagonal
entries of $\mathbf{G}$ are small compared to the diagonal entries,
already gives the general trends, which can be confirmed numerically
for larger off-diagonal entries: When exploitative interaction, where
$G_{ij}$ and $G_{ji}$ ($i\ne j$) have opposite signs, dominate (as for
direct predator-prey interactions), the off-diagonal entries lead to a
narrowing of the eigenvalue distribution along the real axis.  The
general expectation, e.g., with statistically unrelated or positively
correlated $G_{ij}$ and $G_{ji}$, however, is that the distribution
will be broadened, see, e.g. Proposition~1 of
\cite{raghavan06:_stabil_system_random_param}, where a situation
similar to the uncorrelated case is described.  Correspondingly, one
can expect that the real parts of the eigenvalues of $\mathbf{H}$ are
distributed more narrowly than those of $\mathbf{G}$ and so the
condition of $\mathbf{H}$ is improved, except when exploitative
interactions dominate.

The case of the \NP{} objective
($\hat{\mathbf{G}}=\diag(\mathbf{G}^{-1})^{-1}$) is similar to that
for \NS{}, just that now, in addition, the diagonal elements are
modified.  Applying the blockwise matrix inversion formula
\citep{bernstein05:_matrix_mathem} to compute the first diagonal
element of $\mathbf{G}^{-1}$ as a $1\times 1$ block, one obtains the
first diagonal element of $\mathbf{H}$ for \NP{} from that for \NS{}
by adding to it
$-\mathbf{G}_{1,2...S_F} (\mathbf{G}_{2...S_F,2...S_F})^{-1}
\mathbf{G}_{2...S_F,1}/2$,
where we specified matrix blocks by corresponding index ranges.
Corresponding formulae apply for the other diagonal elements of
$\mathbf{H}$.  Sign and magnitude of these additions depends on the
strength and nature of interactions between species, and also on the
condition of the $(S_F-1)\times(S_F-1)$ block inversions.  This
operation tends to broaden the distribution of eigenvalues of
$\mathbf{H}$ along the real axis, so generally worsening the condition
of $\mathbf{H}$ for \NP{} compared to that for \NS{}.

Concluding, these considerations suggest that, for the computation of
$\mathbf{B}_{\text{MSY}}$ and $\mathbf{F}_{\text{MSY}}$ reference
points, structural instability always has a strong impact under the
\TY{} objective, generally a small impact under the \NS{} objective,
and for \NP{} an impact that is typically larger than for \NS{} but
smaller than for \TY{}.

\section{How yields are affected by inaccuracies in mHCR reference
  points}
\label{sec:effect-struct-refpoints}

Above, we discussed in several steps how structural instability of
ecosystems (or operating models simulating these) can lead to
inaccurate values for the reference points ${\mathbf{B}}_{\text{MSY}}$
and ${\mathbf{F}}_{\text{MSY}}$ entering mHCR.  To build intuition as
to what the implications of these inaccuracies are for the resulting
long-term yields, we consider here briefly the hypothetical situation
where the actual ecosystem was describable by a Schaefer model of the
same structure as Eq.~\eqref{eq:surplus-production} of the main text,
with but with true parameters $\underline{\mathbf{r}}$ and
$\underline{\mathbf{G}}$ that generally differ from the ``empirical''
parameters $\mathbf{r}$ and $\mathbf{G}$.  If the mHCR are
successfully implemented and so the respective targets attained, the
resulting total yield from \STC{} will be
\begin{align}
  \label{eq:STC-yield}
  Y={\mathbf{B}}_{\text{MSY}}\tr \left(
    \underline{\mathbf{r}}-\underline{\mathbf{G}}{\mathbf{B}}_{\text{MSY}}
  \right)
\end{align}
and from \PTC{} 
\begin{align}
  \label{eq:PTC-yield}
  Y={\mathbf{F}}_{\text{MSY}}\tr \underline{\mathbf{G}}^{-1}\left(
    {\underline{\mathbf{r}}-\mathbf{F}}_{\text{MSY}}\right).
\end{align}
Similar formulae apply for the individual yield from each stock.

Because in Eq.~\eqref{eq:STC-yield} $\underline{\mathbf{G}}$ enters
directly, small errors in $\mathbf{B}_{\text{MSY}}$ will only have
small effect on yields.  In Eq.~\eqref{eq:PTC-yield}, small errors in
$\mathbf{F}_{\text{MSY}}$ can have large impacts on yields if
$\underline{\mathbf{G}}$ is ill-conditioned and so the system
structurally unstable (which we expect).  Hence \PTC{} is expected to
result in larger deviations from the expected yield than \STC{}.  From
other sources of uncertainty in pressures, such as recruitment
variability or insufficient control by managers of fishing
mortalities, similar adverse effects of structural instability can be
expected.

A more detailed analysis would take account of the error structure in
reference points resulting from the proposed calculation methods.
This analysis is not performed here, because our numerical MSE
provides similar information.

\section{Conservatism through matrix regularisation}
\label{sec:impl-cons-reg}

The manifestation of structural instability discussed in
Section~\ref{sec:effects-struct-inst-par} implies that the
mathematical problem underlying the computation of multispecies MSY
reference points falls into the category of ill-posed problems
\citep{hadamard1923:_illposed}. In practice, such problems are solved
using so-called regularisation methods
\citep{engl2000:_regularization}.  In our case we propose to compute
mHCR parameters from the parameters $\mathbf{r}$ and $\mathbf{G}$ of
the management model not using Eq.~\eqref{eq:BMSY} directly, but
instead as
\begin{align}
  \label{eq:BMSY-reg}
  \begin{split}
    \mathbf{B}_{\text{MSY}}&= \left(
      \mathbf{G}+\hat{\mathbf{G}}\tr+\kappa \mathbf{I} \right)^{-1}
    \left( \mathbf{r}+\kappa \mathbf{B} \right)\\
    & =\left( 2 \mathbf{H}+\kappa \mathbf{I} \right)^{-1} \left(
      \mathbf{r}+\kappa \mathbf{B} \right)
  \end{split}
\end{align} 
and $\mathbf{F}_{\text{MSY}}=\hat{\mathbf{G}}\tr
\mathbf{B}_{\text{MSY}}$, where $\kappa\ge 0$ is a parameter of
dimensions $\text{Time}^{-1}\text{Biomass}^{-1}$ or
$\text{Time}^{-1}$ $(\text{Biomass}/\text{Area})^{-1}$, depending on how
biomass is quantified.

With $\kappa=0$, Eq.~\eqref{eq:BMSY-reg} reduces to the simpler
Eq.~\eqref{eq:BMSY} without conservatism, derived above.  The term
$\kappa I$ in Eq.~\eqref{eq:BMSY-reg} shifts the eigenvalues of
$\mathbf{H}$ away from zero, so improving the condition of the matrix
inversion.  This method is known as Lavrentiev regularisation
\citep{engl2000:_regularization}.  With this modification,
Eq.~\eqref{eq:BMSY-reg} implements a form of conservatism: if $\kappa$
is very large, Eq.~\eqref{eq:BMSY-reg} reduces to
$\mathbf{B}_{\text{MSY}}=\mathbf{B}$, i.e.\ a recommendation to keep
stock sizes at their current levels.  Equation~\eqref{eq:BMSY-reg}
applies conservatism only to those components of $\mathbf{B}$ that are
aligned with eigenvectors of $\mathbf{H}$ corresponding to eigenvalues
near zero, i.e.\ where conservatism is most useful because numerical
uncertainty is large.  This distinguishes Eq.~\eqref{eq:BMSY-reg} from
alternative conceivable formulae, e.g.
$ \mathbf{B}_{\text{MSY}}= (1+\kappa)^{-1}[\mathbf{H}^{-1}
\mathbf{r}/2+\kappa \mathbf{B}] $,
which is also ``conservative'' for large $\kappa$.  Conservatism is
expected to be beneficial if $\mathbf{H}$ is ill-conditioned, and
otherwise to be neutral or detrimental in its effect on the ability of
managers to meet objectives.

One might expect that the matrix inversion involved in computing
$\hat{\mathbf{G}}=\diag(\mathbf{G}^{-1})^{-1}$, required for
management towards \NP{} objectives, could also be made more robust
using regularisation methods, but we did not find a regularisation
schemes that would improve outcomes among the schemes we experimented
with.  Except for this, \SOC{} does not require any matrix inversion,
so the above considerations play no role if \SOC{} is employed.

\section{Summary of theoretical expectations}
\label{sec:summ-theor-expect}

Based on the preceding discussions, one can expect that, among the
objectives we considered, conservatism will be most important for
achieving \TY{} and least important for \NS{}, while for achievement
of the \NP{} objective conservatism has an intermediate role to play
(Section~\ref{sec:effects-struct-inst-par}).  On the other hand, it is
clear that too strong conservatism will unnecessarily increase the
number of adaptive management cycles required to approach management
goals.  Among the target-oriented management strategies, \STC{} is
expected to be more robust than \PTC{}
(Section~\ref{sec:effect-struct-refpoints}).  For \SOC{} strategies,
conservatism is unnecessary (Section~\ref{sec:impl-cons-reg}).

We note that, based on theoretical considerations alone, it is
difficult to formulate general expectations of whether total yields
will be higher under the \NP{} or the \NS{} objective.  Consider, for
example, the case $S_F=2$, where, based on above calculations and
disregarding model uncertainty, the difference between yields under
the two objectives is
\begin{align}
  \label{eq:yield-difference}
  Y_{\text{NS}}-Y_{\text{NP}}=\frac{G_{12}G_{21}(G_{12}+G_{21})(2G_{11}G_{22}r_1r_2-G_{21}G_{22}r_1^2-G_{11}G_{12}r_2^2)}{(G_{11}G_{22}-G_{12}G_{21})(4G_{11}G_{22}-G_{12}G_{21})^2}.
\end{align}
While for perfectly symmetric competition ($G_{11}=G_{22}$,
$G_{12}=G_{21}$, $r_1=r_2$) this difference evaluates to
$4 G_{11}G_{12}^2r_1^2(4G_{11}^2-G_{12}^2)^{-2}(G_{11}+G_{12})^{-1}$,
which is typically positive, it is clear that the sign of
Eq.~\eqref{eq:yield-difference} will flip, e.g., whenever either that
of $G_{12}$ or $G_{21}$ flips.  In general, pursuit of either \NP{} or
\NS{} objective could therefore result in higher total yields.

\section{Detailed quantitative outcomes of the MSE}
\label{sec:deta-quant-outc}

We now present details of the numerical MSEs we conducted based on two
sets of communities, generated by using the PDMM with the scaling
parameter set to $X=1$ and $X=2$, respectively
(Section~\ref{sec:model-struct-param}).
\begin{figure}[htbp]

  \begin{minipage}{0.45\linewidth} 
    \includegraphics[width=\linewidth]{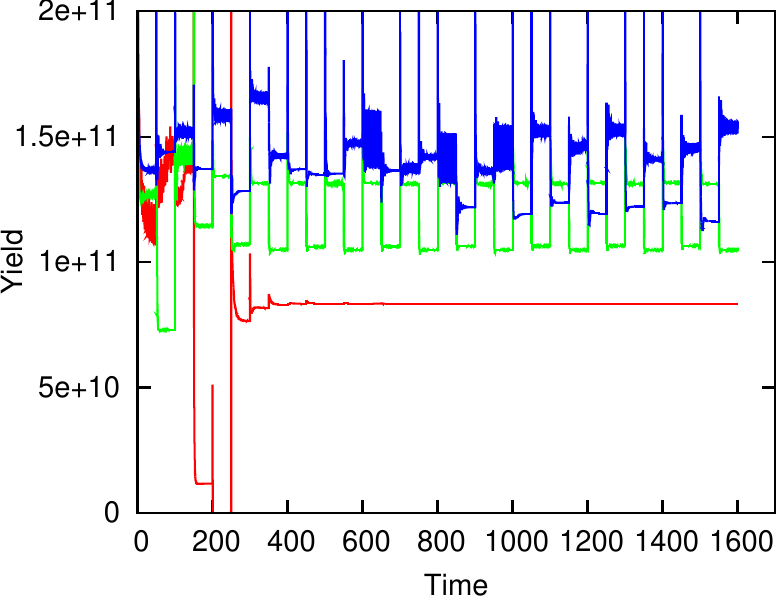}
  \end{minipage}  
  \begin{minipage}{0.45\linewidth} 
    \includegraphics[width=\linewidth]{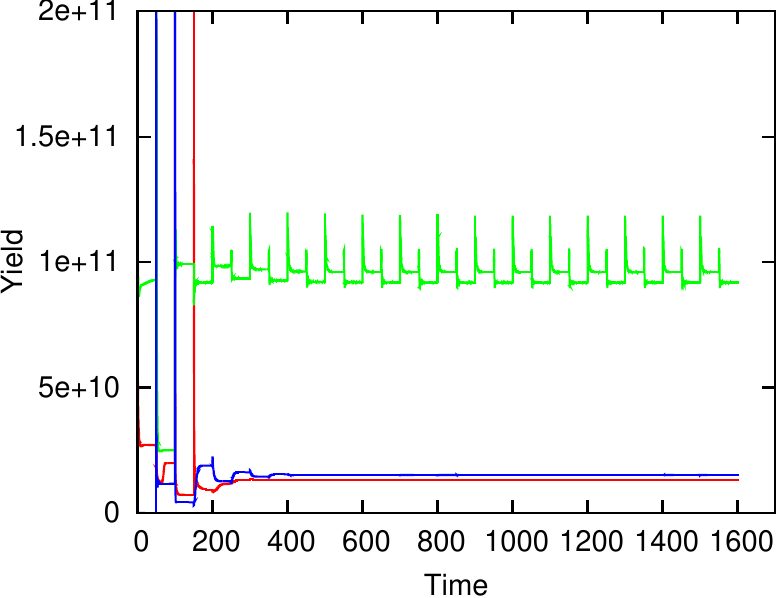}
  \end{minipage}  

  \begin{minipage}{0.49\textwidth} \centering (a)
  \end{minipage}  
  \begin{minipage}{0.49\textwidth} \centering (b)
    
  \end{minipage}  

  \begin{minipage}{0.45\linewidth} 
    \includegraphics[width=\linewidth]{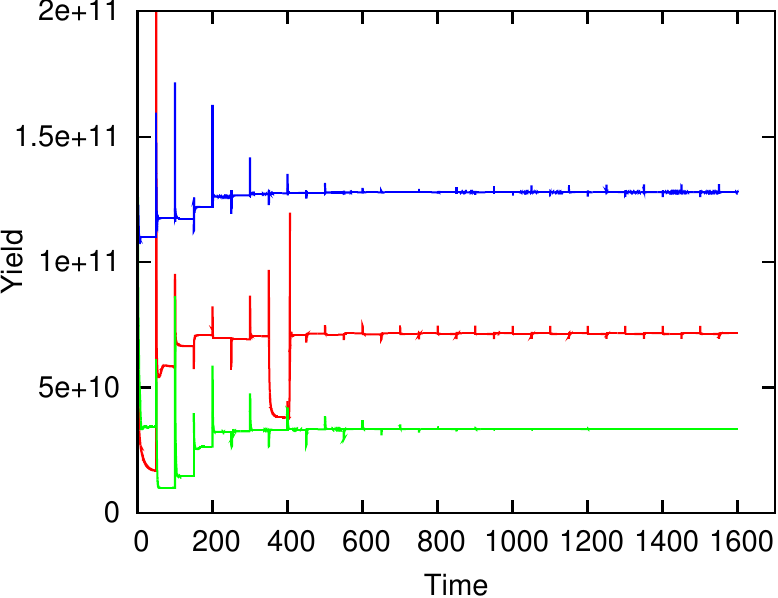}
  \end{minipage}  
  \begin{minipage}{0.45\linewidth} 
    \includegraphics[width=\linewidth]{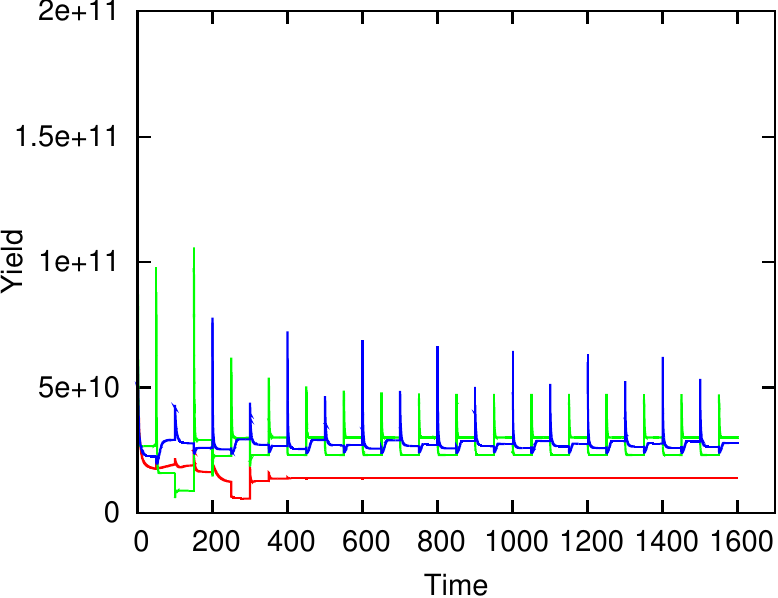}
  \end{minipage}  

  \begin{minipage}{0.49\textwidth} \centering (c)
  \end{minipage}  
  \begin{minipage}{0.49\textwidth} \centering (d)
  \end{minipage}  

  \caption{Yield time series when applying \PTC{} (red), \STC{}
    (green) and \SOC{} (blue) strategies for the Total Yield objective
    on the $X=2$ sample communities saved after (a) $6\cdot 10^4$, (b)
    $10^5$, (c) $1.5\cdot 10^5$ and (d) $2\cdot 10^5$ species added.}
  \label{y4webs}
\end{figure}

\subsection{Variability among MSE runs}
\label{sec:vari-among-mse}

Figure~\ref{y4webs} shows typical time series of simulated
yields. They correspond to applying the \PTC{}, \STC{}, and \SOC{}
management strategies for the \TY{} objective over a period of 32
management cycles to four of the sample communities ($X=2$) that we
studied.  It is apparent that yields can vary substantially between
management strategies and between sample communities, and that even
the ranking of yields achieved using the three strategies can vary.
In order to establish the relative performance of different management
plans we therefore tested them on a larger number of sample
communities.

\subsection{Standard regularisation and optimal regularisation}
\label{ssec:reg} 

The numerical results for plans involving a ``standard'' degree of
conservatism were obtained using a fixed regularisation parameter
$\kappa=5\,\mathrm{g}^{-1}\mathrm{m}^{2}\mathrm{yr}^{-1}$ in all cases
considered and for all sample communities.  To put this value into
context, we computed the ratios $F_{\text{MSY}}/B_{\text{MSY}}$ for
\CFP{} for all fish species from all virgin model communities, and
found that our choice for $\kappa$ corresponds approximately to the
$2/3$ quantile of their distribution.  This comparison should be
useful for choosing a corresponding regularisation parameter in
practical applications.  In particular, it is noteworthy that $\kappa$
is not such a ``small'' parameter here as the theory of matrix
regularisation might suggest.



\begin{figure}

  \vspace*{-45mm}
  
  \hspace*{-15mm}
  \begin{minipage}{0.45\linewidth} 
    \centering
    \includegraphics[angle=0,scale=0.5]{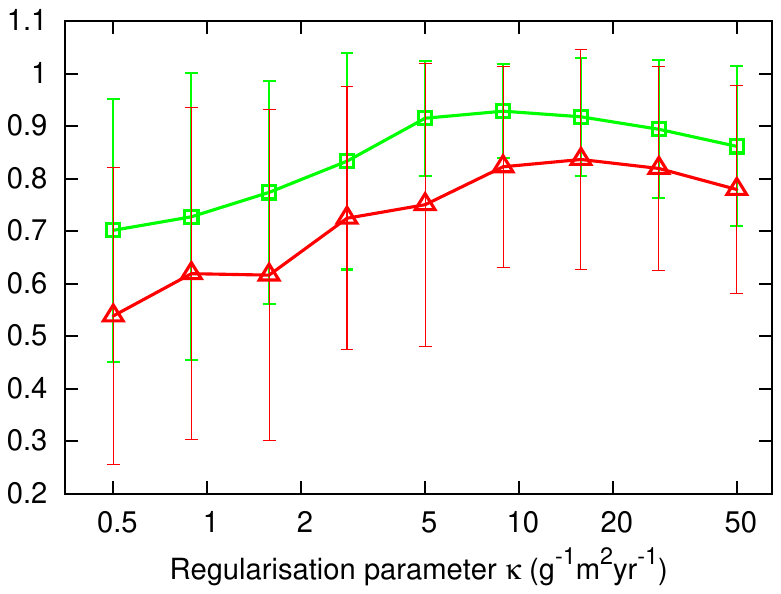}
  \end{minipage}  \hspace*{10mm}
  \begin{minipage}{0.45\linewidth} 
    \centering
    \includegraphics[angle=0,scale=0.5]{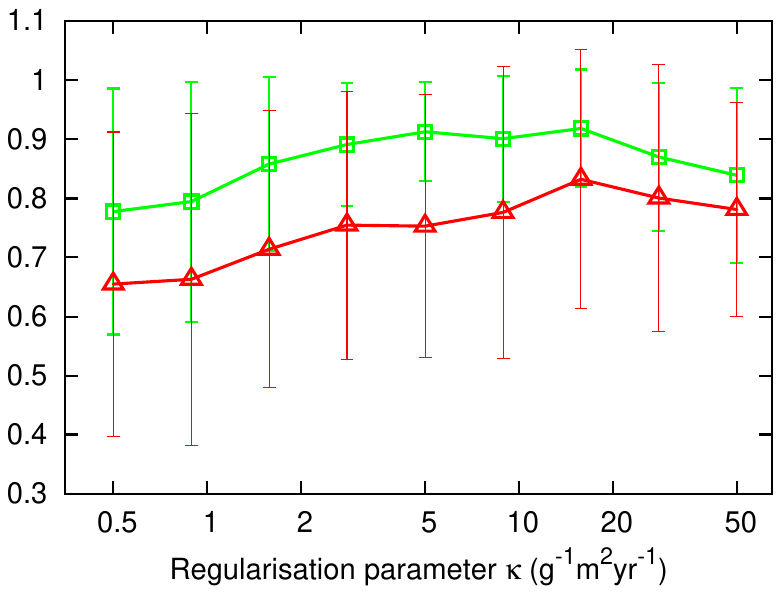}
  \end{minipage}  \vspace*{-10mm}

  \begin{minipage}{0.49\textwidth} \centering (a)(i)
  \end{minipage}  \hspace*{0mm}
  \begin{minipage}{0.49\textwidth} \centering (b)(i)
    
  \end{minipage}  \vspace*{-40mm}

  \hspace*{-15mm}
  \begin{minipage}{0.45\linewidth} 
    \centering
    \includegraphics[angle=0,scale=0.5]{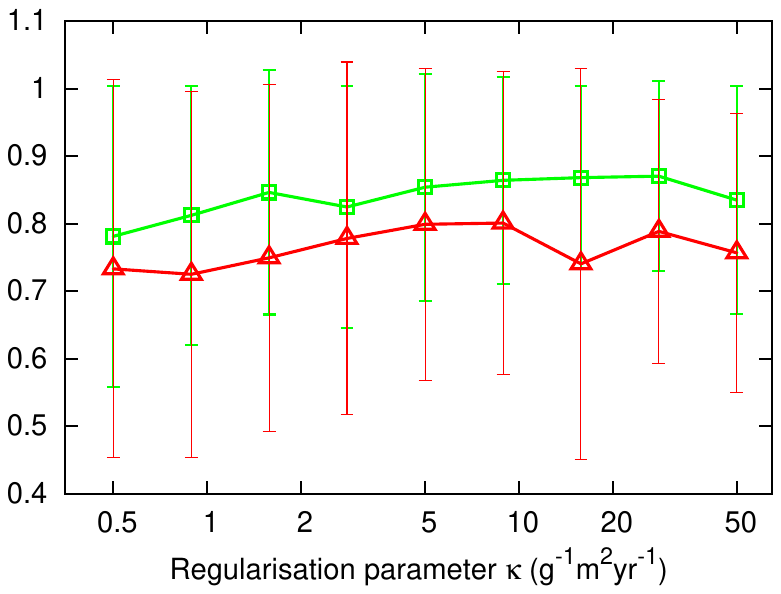}
  \end{minipage}  \hspace*{10mm}
  \begin{minipage}{0.45\linewidth} 
    \centering
    \includegraphics[angle=0,scale=0.5]{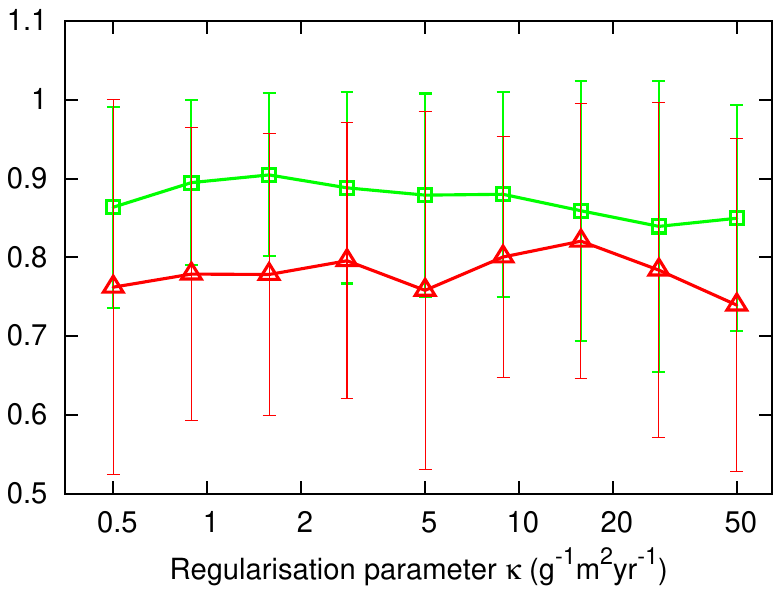}
  \end{minipage}  \vspace*{-10mm}

  \begin{minipage}{0.49\textwidth} \centering (a)(ii)
  \end{minipage}  \hspace*{0mm}
  \begin{minipage}{0.49\textwidth} \centering (b)(ii)
  \end{minipage}  \vspace*{-0mm}

  \caption{Dependence on the regularisation parameter $\kappa$ of the
    mean (lines and markers) and standard deviation (bars) of the
    normalised yield calculated across all (a) $X=1$ and (b) $X=2$
    sample communities, when managing towards (i) \TY{} and (ii) \NP{}
    objectives. The results for \PTC{} are shown with red lines and
    triangle markers, those for \STC{} with green lines and square
    markers.}
  \label{varreg}
\end{figure}

To compare this choice of $\kappa$ with outcomes for other choices, we
conducted, for each plan and each sample community, a series of 9 MSE,
using a set of 9 log-equidistant values of the regularisation
parameter ranging from
$\kappa=0.5\,\mathrm{g}^{-1}\mathrm{m}^{2}\mathrm{yr}^{-1}$ to
$\kappa=50\,\mathrm{g}^{-1}\mathrm{m}^{2}\mathrm{yr}^{-1}$.  To
facilitate the comparison, we normalised for each sample community and
each plan the average yields to the maximum value across the 9 runs,
and from this computed mean and standard deviations across all sample
communities. Thus, the value $1$ for the mean (and standard deviation
$0$) associated with a certain regularisation parameter value would
indicate that the use of that value produces always (i.e for all
sample communities) the highest yield for the plan under
consideration.  The calculated means and standard deviations for both
$X=1$ and $X=2$ communities are shown in Figure~\ref{varreg} (a) and
(b), respectively.

The results of Figure \ref{varreg} show that our choice of the
standard value
$\kappa=5\,\mathrm{g}^{-1}\mathrm{m}^{2}\mathrm{yr}^{-1}$ generally
performs close to the ``best'' value (i.e. the value corresponding to
the largest means shown in Figure \ref{varreg}).  While some modest
yield improvements can be achieved in some cases by adjusting
$\kappa$, it should be observed that the standard deviations shown in
Figure \ref{varreg} remain large even for the points with the highest
mean (i.e. the ``best'' $\kappa$), indicating high variability among
sample communities in terms of the best-performing $\kappa$.  This
idiosyncratic behaviour appears to be most pronounced for \PTC{} and
for Nash-type objectives.

\subsection{Detailed quantitative results}
\label{sec:deta-quant-outc-1}

\begin{table}
  \caption{Key to nomenclature used in Figures~\ref{verticalX1X2} 
    and~\ref{ttestX2}}\label{tab:key}
  \centering
  \begin{tabular}{|cccc|}
    \hline
    optimal &conservatism&
    \begin{minipage}[c]{\widthof{\NP}}
      \centering\NP\\\NS\\\TY
    \end{minipage}
    &
    \begin{minipage}[c]{\widthof{\PTC\newline\SOC}}
      \vspace{0.2em}\centering\PTC\\\STC\\\SOC\vspace{0.2em}
    \end{minipage}
    \\
    \hline
    o&c&  \begin{minipage}[c]{\widthof{\NP}}
      \centering Np\\Ns\\-
    \end{minipage}
    &  
    \begin{minipage}[c]{\widthof{\PTC\newline\SOC}}
      \vspace{0.2em}\centering PTC\\STC\\SOC\vspace{0.2em}
    \end{minipage}
    \\
    \hline
    \hline
    \multicolumn{4}{|c|}{SSC: \CFP\quad MSYt: theoretical
      MSY}    
    \\
    \hline
\end{tabular}

\end{table}

\begin{figure}

  \hspace*{\fill}
  \begin{minipage}{0.4\linewidth} 
    \begin{center}
      \qquad \qquad $\mathsf{\mathit{X}=1}$\vspace{-1.5em}
    \end{center}
    \includegraphics[angle=270,width=\textwidth]{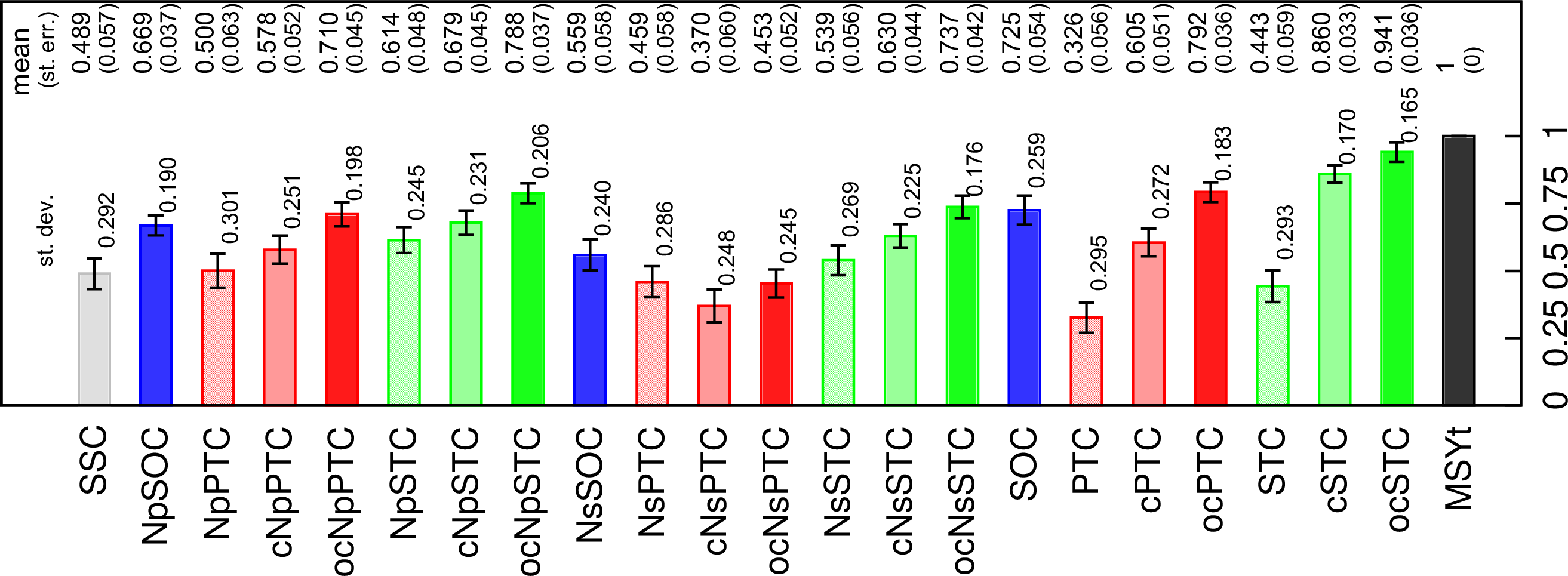}
  \end{minipage} 
  \hspace*{\fill}
  \begin{minipage}{0.4\linewidth} 
    \begin{center}
      \qquad \qquad $\mathsf{\mathit{X}=2}$\vspace{-1.5em}
    \end{center}
    \includegraphics[angle=270,width=\textwidth]{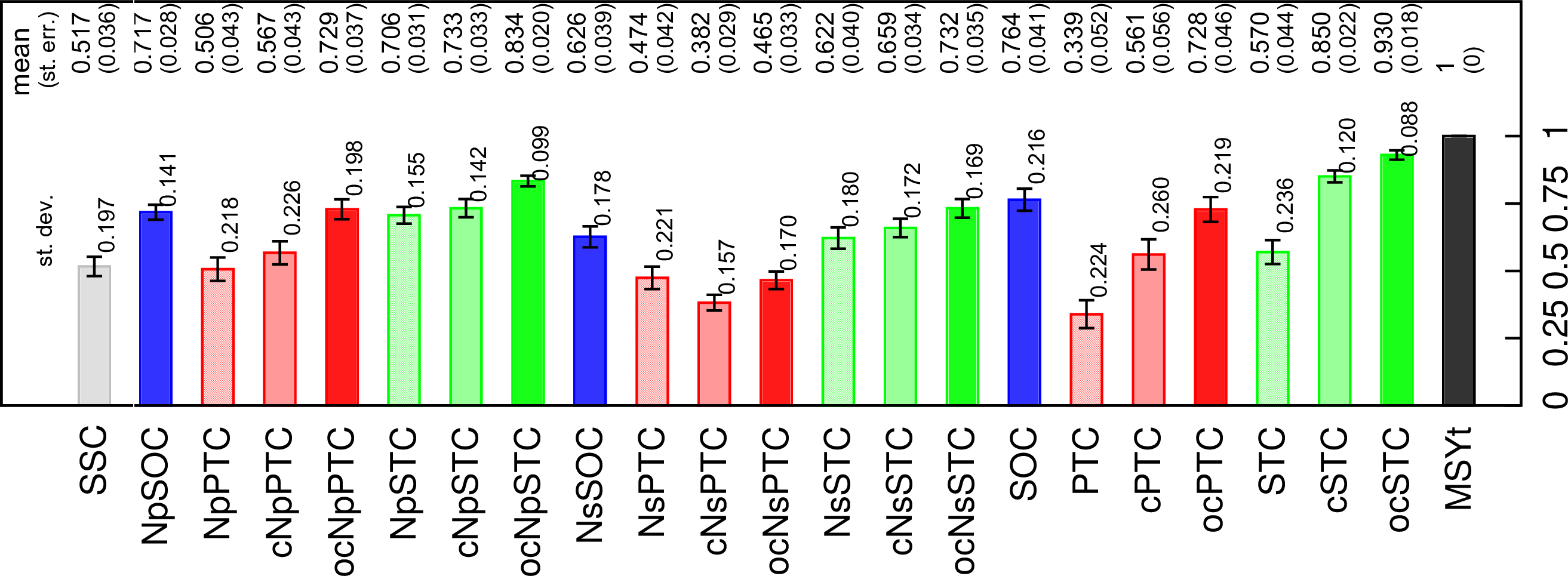}
  \end{minipage} 
  \hspace*{\fill}

  \caption{The means (bars), standard deviations and the standard
    errors of the mean (error bars) for total yields (in proportion to
    theoretical maximum MSYt), calculated across the all $X=1$ (left)
    $X=2$ (right) model communities, when applying \SOC{} (blue),
    \PTC{} (red) and \STC{} (green) strategies, compared with results
    for \CFP{} (grey).}
  \label{verticalX1X2}
\end{figure}

%
\begin{figure}
  \begin{center}
    \small \textsf{Comparison of yields}
  \end{center}
  \vspace{-0.3cm}
  \includegraphics[width=1.0\textwidth]{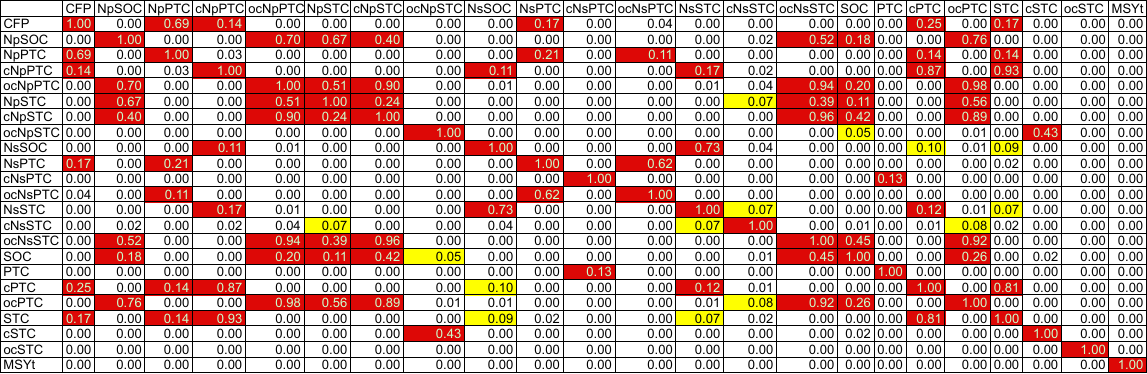}
  \begin{center}
    \small \textsf{Comparison of survival rates}
  \end{center}
  \vspace{-0.3cm}
  \includegraphics[width=1.0\textwidth]{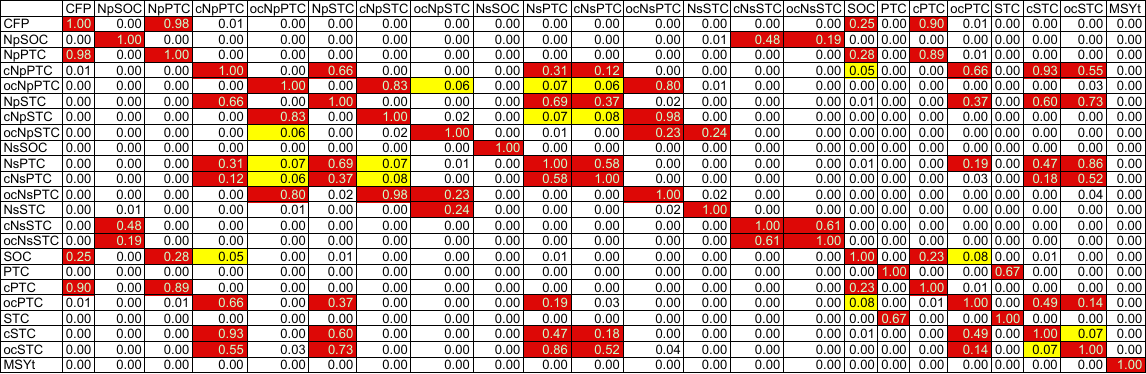}
  \begin{center}
    \small \textsf{Comparison of mean log sizes}
  \end{center}
  \vspace{-0.3cm}
  \includegraphics[width=1.0\textwidth]{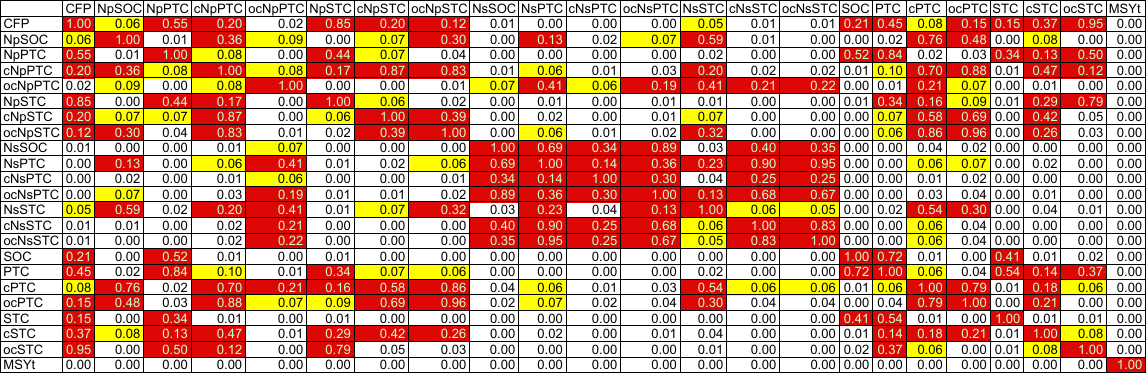}
  
  \caption{The $p$-values from two-sided paired t-tests comparing
    normalised yields (top), survival rates (centre) and mean log
    sizes (bottom) for each pairing of 23 plans over the 37 sample
    communities for $X=2$, corresponding to Figure~\ref{verticalX1X2}
    (right).  Values larger than 0.05 and 0.1 are highlighted in
    yellow and red, respectively.}
  \label{ttestX2}
\end{figure}

In Figures~\ref{verticalX1X2} and~\ref{ttestX2} we display a detailed
quantitative comparison of the outcomes of the MSEs.  A key to the
codes we used for different plans is given in Table~\ref{tab:key}.
Figure~\ref{verticalX1X2} displays graphically and numerically mean
and standard deviation of the proportion of theoretical MSY attained,
as well as the standard error of the mean.  In order to account for a
weak correlation between results from subsequently sampled
communities, the standard errors of the mean in
Figure~\ref{verticalX1X2} were computed based on the ``effective''
number of degrees of freedom following
\cite{pyperpeterman98:_correlation}, rather than the actual sample
size (total number of sample communities).

Because all plans were evaluated based on the same set of sample
communities, so that MSE were not statistically independent, we used
two-sided paired t-tests to assess the statistical significance of
differences between outcomes of different plans.  Outcomes of the
t-tests for $X=2$ for each pair of plans are shown in
Figure~\ref{ttestX2}.  For the smaller ($X=1$) communities, this
comparison had slightly lower statistical power, but otherwise
produced very similar results.


\putbib{}
\end{bibunit}



\end{document}